\documentclass[iop,numberedappendix,appendixfloats]{emulateapj}
\usepackage{epstopdf}
\usepackage{graphics}
\usepackage{natbib}

\newcommand{\msun}{\mbox{$\rm M_{\odot}$}}

\newcommand{\kms}{\mbox{km s$^{-1}$}}

\newcommand{\etal}[1]{{ et al.}~}
\def\kms{\ifmmode \hbox{km~s}^{-1}\else km~s$^{-1}$\fi}
\def\etal {{\it et al.}}
\def\deg      {{\ifmmode^\circ\else$^\circ$\fi} } %%% Overwrites TeX \deg

\def\h2     {H$_2$}

\def\h2     {H$_2$}

\def\arcsec{\hbox{$^{\prime\prime}$}}

\shorttitle{ISM Masses and SF Law}
\shortauthors{Scoville \etal}

\begin{document}

\title{ISM Masses and the Star Formation Law at z = 1 to 6   \\ ALMA Observations of Dust Continuum in 145 Galaxies \\ in the COSMOS Survey Field}
 \author{ N. Scoville\altaffilmark{1},  K. Sheth\altaffilmark{3}, H. Aussel\altaffilmark{2}, P. Vanden Bout\altaffilmark{10}, P. Capak\altaffilmark{6}, A. Bongiorno\altaffilmark{12},  C. M. Casey\altaffilmark{11}, L. Murchikova\altaffilmark{1}, J. Koda\altaffilmark{13},  J. \'Alvarez-M\'arquez\altaffilmark{14}, N. Lee\altaffilmark{5}, C. Laigle\altaffilmark{15}, H. J. McCracken\altaffilmark{15}, O. Ilbert\altaffilmark{14}, A. Pope\altaffilmark{8},  D. Sanders\altaffilmark{5}, J. Chu\altaffilmark{5}, S. Toft\altaffilmark{7}, R.\,J. Ivison\altaffilmark{9,16} and S. Manohar\altaffilmark{1}}

\altaffiltext{1}{California Institute of Technology, MC 249-17, 1200 East California Boulevard, Pasadena, CA 91125}
\altaffiltext{2}{AIM Unit\'e Mixte de Recherche CEA CNRS, Universit\'e Paris VII UMR n158, Paris, France}
\altaffiltext{3}{North American ALMA Science Center, National Radio Astronomy Observatory, 520 Edgemont Road, Charlottesville, VA 22901, USA}
%\altaffiltext{4}{National Optical Astronomy Observatory, 950 North Cherry Avenue, Tucson, AZ 85719, USA}
\altaffiltext{5}{Institute for Astronomy, 2680 Woodlawn Dr., University of Hawaii, Honolulu, Hawaii, 96822}
\altaffiltext{6}{Spitzer Science Center, MS 314-6, California Institute of Technology, Pasadena, CA 91125}
\altaffiltext{7}{AD(Dark Cosmology Centre, Niels Bohr Institute, University of Copenhagen, Juliana Mariesvej 30, DK-2100 Copenhagen, Denmark)}
\altaffiltext{8}{Department of Astronomy, University of Massachusetts, Amherst, MA 01003}
\altaffiltext{9}{Institute for Astronomy, University of Edinburgh, Blackford Hill, Edinburgh EH9 3HJ, UK}
\altaffiltext{10}{National Radio Astronomy Observatory, 520 Edgemont Road, Charlottesville, VA 22901, USA}
\altaffiltext{11}{Department of Astronomy, The University of Texas at Austin, 2515 Speedway Blvd Stop C1400, Austin, TX 78712}
\altaffiltext{12}{INAF - Osservatorio Astronomico di Roma, Via di Frascati 33, I-00040 Monteporzio Catone, Rome, Italy}
\altaffiltext{13}{Department of Physics and Astronomy, SUNY Stony Brook, Stony Brook, NY 11794-3800, USA}
\altaffiltext{14}{Laboratoire dÕAstrophysique de MarseilleÑLAM, UniversitŽ dÕAix-Marseille \& CNRS, UMR7326, 38 rue F. Joliot-Curie, F-13388 Marseille Cedex 13, France}
\altaffiltext{15}{ CNRS, UMR 7095, Institut dÕAstrophysique de Paris, F- 75014, Paris, France}
\altaffiltext{16}{ESO, Karl-Schwarzschild-Strasse 2, D-85748 Garching, Germany}

%\maketitle

%~~~~~~~~~~~~~~~~~~~~~~~~~~~~~~~~~~~~~~~~~~~~Accepted ApJ 2/25/13

\altaffiltext{}{}

\begin{abstract} 
ALMA Cycle 2 observations of the long wavelength dust emission in 145 star-forming galaxies are used to probe the evolution 
of star-forming ISM. We also develop the physical basis and empirical calibration (with 72 low-z and z $\sim$2 galaxies) for using the 
dust continuum as a quantitative probe of interstellar medium (ISM) masses. 
The galaxies with highest star formation rates (SFRs) at $\rm <z>$ = 2.2 and 4.4 have gas masses up to 100 times that of the Milky Way and gas mass fractions reaching 50 to 80\%, i.e. gas masses 1 - 4 $\times$ their stellar masses. 
 We find a single high-z star formation law: $\rm SFR = 35~ M_{\rm mol}^{0.89} \times (1+z)_{z=2}^{0.95} \times (sSFR)_{MS}^{0.23}$ \msun yr$^{-1}$ -- an {\bf approximately linear
dependence on the ISM mass and an increased star formation efficiency per unit gas mass at higher redshift}. Galaxies above the Main Sequence (MS) have larger gas masses but are converting their ISM into stars on a timescale only slightly shorter than those on the MS -- thus these 'starbursts' are largely the result of having greatly increased gas masses rather than and increased efficiency for converting gas to stars.  At z $> 1$, the entire population of star-forming galaxies has $\sim$ 2 - 5 times shorter gas depletion times than low-z galaxies. These {\bf shorter depletion times indicate a different mode of star formation in the early universe} --  most likely dynamically driven by compressive, high-dispersion gas motions -- a natural consequence of the high gas accretion rates.
\end{abstract}

\medskip

 \keywords{cosmology: observations --- galaxies: evolution --- galaxies: ISM}

\section{Introduction}\label{intro}
For star forming galaxies there exists a Main Sequence (MS) with galaxy star formation rates (SFRs) varying nearly linearly with stellar mass \citep{noe07}.  At z  $\sim$ 2 approximately 2\% of the star forming galaxies have SFRs more than 4 times higher than the MS, contributing $\sim10$\% of the total star formation \citep{rod11}. These are often identified as the starburst galaxy population \citep[][]{elb11,sar12}.

The specific star formation rate, (sSFR $\equiv$ SFR/M$_{stellar}$), is roughly constant along the MS at each cosmic epoch but increases 20-fold out to z $\sim 2.5$, consistent with  the overall increase in the cosmic star formation rate \citep{hop06,kar11,whi12,lee15}. Understanding the cause of the MS evolution 
and the nature of galaxies above the MS is fundamental to understanding the cosmic evolution of star formation. 

The interstellar medium (ISM) fuels the activities of both galactic star formation and galactic nuclei -- in both cases, peaking at z $\sim$ 2.
Is the cosmic evolution of these activities simply due to galaxies having larger ISM masses (M$_{\rm ISM}$) at earlier epochs, or are they forming stars with a higher efficiency ($\epsilon \equiv 1/\tau_{SF}$ = SFR/M$_{\rm ISM}$)?
Specifically: 1) is the 20-fold increase in the SFR of the MS from z = 0 to 2 due to proportionally increased gas contents at early epochs 
or due to a higher frequency of starburst activity? 
and 2) are galaxies above the MS converting their gas to stars 
with higher efficiency or do they simply have more gas than those on the MS? Measurements of galactic ISM gas contents are critical to answering these very basic questions.

Over the last decade, the rotational transitions of CO  have been used to probe the molecular ISM of high redshift galaxies \citep{sol05,cop09,tac10,cas11,bot13,tac13,car13}. 
Here, we employ an alternative approach -- using the long wavelength dust continuum to probe ISM masses, specifically, \it molecular \rm ISM masses, at high 
redshift \citep{sco12,eal12,mag12}. This dust emission is optically thin. % and avoids the quandary of the variable CO-to-H$_2$ conversion factor. 
For high stellar mass star-forming galaxies, the dust continuum can also be detected by ALMA in just a few minutes of observing, whereas for the same galaxies, CO would require an hour or more with ALMA. 

\begin{figure*}[ht]
\epsscale{1}  
\plotone{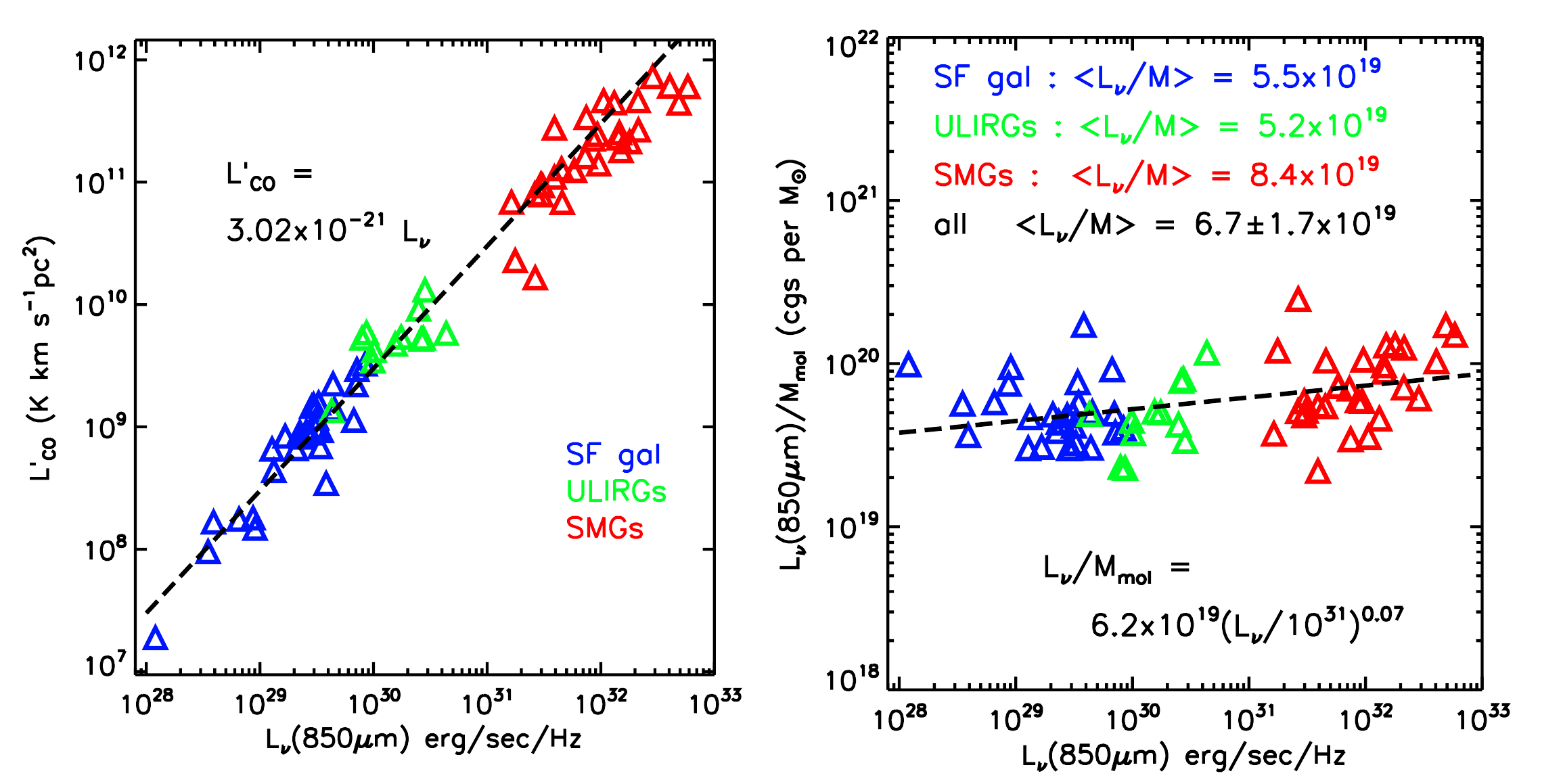}
\caption{{\bf Left:} The CO(1-0) luminosity and $L_{\nu}$ at 850$\mu$m to $M_{\rm mol}$ are shown for three samples 
of galaxies -- normal low-z star forming galaxies, low-z ULIRGs and z $\sim$ 2 SMGs. All galaxies were selected to have global measurements of CO (1-0) and Rayleigh-Jeans dust continuum fluxes. The large range in apparent luminosities is enhanced by including the high redshift SMGs, many of which 
in this sample are strongly lensed. {\bf Right:} The ratio of $ L_{\nu}$ at 850$\mu$m to $M_{\rm mol}$ is shown for the three samples of galaxies, indicating a very similar proportionality constant 
between the dust continuum flux and the molecular masses derived from CO(1-0) emission. The molecular masses were estimated from the CO (1-0) luminosities using a single standard Galactic  $X_{CO} = 3\times10^{20}$ N(H$_2$) cm$^{-2}$ (K km s$^{-1})^{-1}$ (see Appendix \ref{dust_app}). }
\label{empir_cal} 
\end{figure*}

Here, we use a sample of 70 galaxies (28 local star forming galaxies, 12 low redshift ULIRGs and 30 SMGs at z $\sim$ 2) to \emph{empirically calibrate} the ratio of 
long wavelength dust emission to CO (1-0) line luminosity and, hence, molecular gas mass. Without making any corrections for variable dust temperatures, galaxy metallicity or 
CO excitation variations, the ratio of long wavelength dust luminosity to CO (1-0) luminosity and molecular gas mass is found to be remarkably 
constant across this sample, which includes normal star forming galaxies, starburst galaxies at low redshift and massive sub millimeter galaxies (SMGs) at z = 2 - 3 (see Figure \ref{empir_cal}). 

It is particularly significant that the local ULIRGs exhibit the same proportionality between the long wavelength dust continuum and the CO (1-0) luminosity. (This 
argues against a different CO-to-H2 conversion factor in ULIRGs since the physical dependences (on density and $T_D$ or $T_K$) of the mass to dust  and CO emission fluxes
are different -- to be discussed in a future work). 

In both the calibration samples and the sample of high redshift galaxies we observed here with ALMA, we have intentionally restricted the samples to objects with high 
stellar mass ($M_{stellar} = 2\times10^{10} - 4\times10^{11}$\msun); thus we are not probing lower metallicity systems where the dust-to-gas abundance ratio is likely to drop 
or where there could be significant molecular gas without CO \citep[see][]{bol13}. 

\begin{figure}[ht]
\epsscale{1}  
\plotone{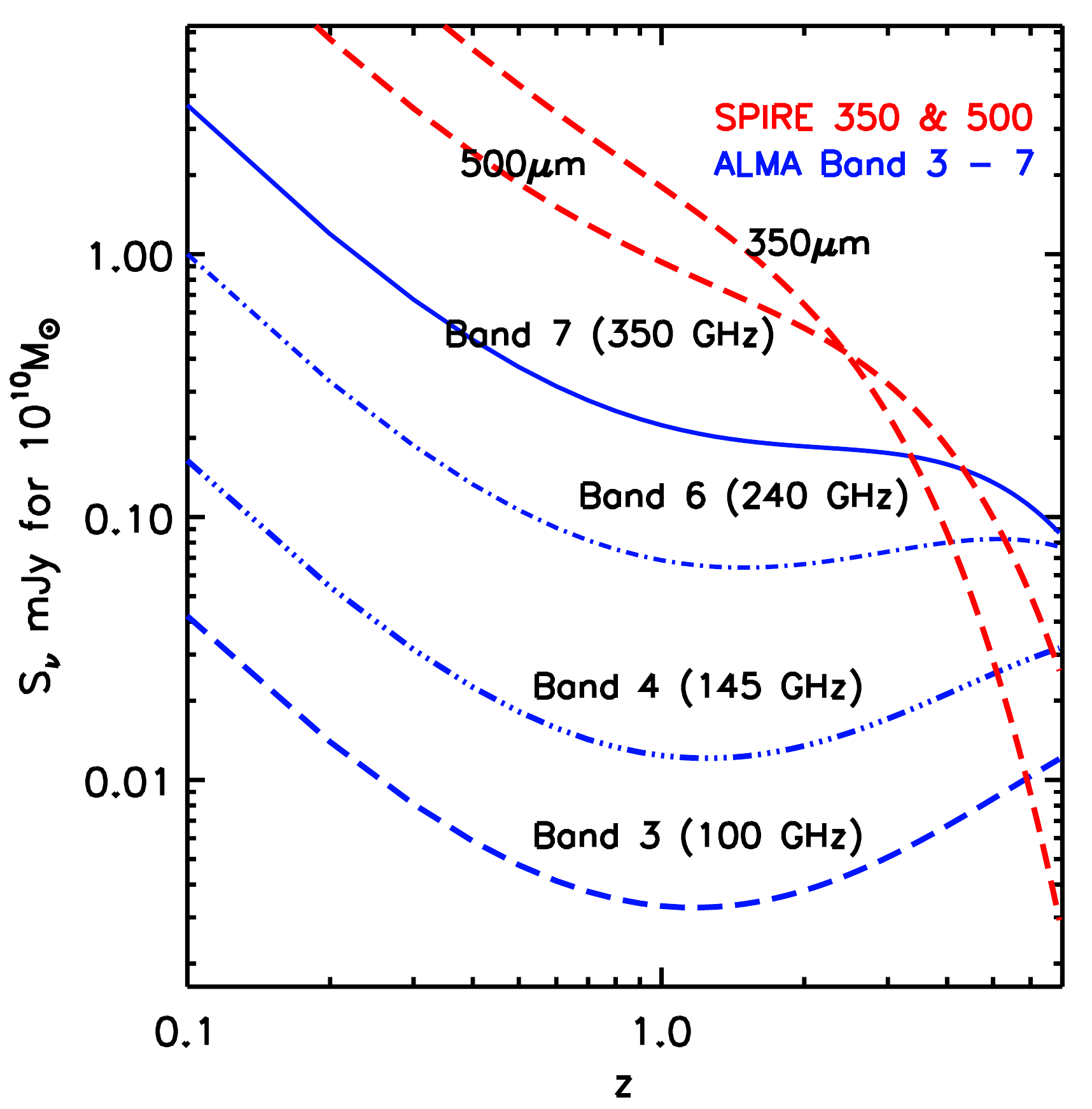}
\caption{The expected continuum fluxes for the ALMA bands at 100, 145, 240 and 350 GHz and for SPIRE 350 and 500$\mu$m for M$_{mol} = 10^{10}$\msun ~derived using the empirical calibration embodied in Equation \ref{alpha}, 
an emissivity power law index $\beta = 1.8$ and including the RJ departure coefficient $\Gamma_{RJ} (25K)$. Since the point source sensitivities of ALMA 
in the 3 bands are quite similar, the optimum strategy is to use Band 7 out to z $\sim 2 -3$; at higher redshift, lower frequency ALMA bands are required to avoid large uncertainties in the RJ correction.}
\label{alma_obs} 
\end{figure}

The sample of galaxies at high redshift observed with ALMA is described in Section \ref{sample}. The stellar mass, SFR, submm flux and estimated gas mass for each individual galaxy are tabulated in the tables in Appendix \ref{source_app}.
Average flux measurements for subsamples of galaxies are presented in Section \ref{obs} and the derived gas masses and gas mass fractions 
in Section \ref{mass}. The implications for the evolution of ISM and star formation at the peak of cosmic activity are discussed in Section \ref{discuss}.

\section{Long Wavelength Dust Continuum as a Gas Mass Tracer}\label{basis}

At long wavelengths, the dust emission is optically thin and the observed flux density is proportional to the mass of dust, the dust opacity coefficient and the 
mean temperature of dust contributing emission at these wavelengths. Here we briefly summarize the foundation for using the dust continuum as a quantitative probe of ISM masses; 
in Appendix \ref{dust_app} we provide a thorough exposition.  

To obviate the need to know explicitly the dust opacity and the dust-to-gas abundance ratio, we 
empirically calibrate  
the ratio of the specific luminosity at rest frame 850$\mu$m to ISM molecular gas mass using samples of observed galaxies -- thus absorbing the opacity curve, abundance ratio
and dust temperature into a single empirical constant $ \alpha_{850\mu \rm m} =  L_{\nu_{850\mu \rm m}} / M_{\rm mol} $.
This procedure was initially done by \cite{sco13} with three galaxy samples.

In Appendix \ref{dust_app}, we have redone the calibration of the mass determination from the submm-wavelength dust continuum. The sample of calibration galaxies is greatly extended and we use Herschel SPIRE 500$\mu$m imaging instead of SCUBA 850$\mu$m. The SPIRE observations recover more accurately the extended flux components of nearby galaxies than the SCUBA 
measurements which were used in \cite{sco13}. (SCUBA observations use beam chopping to remove sky backgrounds and this can compromise the extended flux components.) For the molecular masses, we use CO(1-0) data which is homogeneously calibrated for the local 
galaxies. The empirical calibration samples now consist of 28 local star-forming galaxies (Table \ref{tab:local_gal}), 12 low-z ULIRGs (Table \ref{tab:ulirg}) and 30 z = 1.4 -- 3 SMGs (Table \ref{tab:smg}).

All three samples exhibit the same linear correlation between CO(1-0) luminosity $L'_{CO}$ and $L_{\nu_{850\mu \rm m}}$ as shown in Figure \ref{empir_cal}-Left. To convert the CO luminosities 
to molecular gas masses, we then use a single CO-to-H$_2$ conversion constant for all objects (Galactic: $X_{CO} = 3\times10^{20}$ N(H$_2$) cm$^{-2}$ (K km s$^{-1})^{-1}$) and the 
resultant masses are shown rated to $L_{\nu_{850\mu \rm m}}$ in Figure \ref{empir_cal}-Right. We find a single calibration constant  $\alpha_{850\mu \rm m} $= 6.7
$\times10^{19}  \rm ~ergs~ sec^{-1} Hz^{-1} \msun^{-1}$ (Equation \ref{alpha}). [This value for $\alpha_{850\mu \rm m} $ is in excellent agreement with that obtained from Planck data for the Galaxy
(6.2$\times10^{19}  \rm ~ergs~ sec^{-1} Hz^{-1} \msun^{-1}$, see Section \ref{pla}). The earlier value used by \cite{sco13} was 1$\times10^{20}  \rm ~ergs~ sec^{-1} Hz^{-1} \msun^{-1}$.]

The long wavelength dust emissivity index which is needed to translate observations at different rest frame wavelengths is taken to be $\beta = 1.8 \pm 0.1$, based on the extensive Planck data in the Galaxy \citep{pla11b}. The mass of molecular gas is then derived from the observed flux density using Equation \ref{mass_eq}, which gives the expected flux density at observed frequency $\nu_{obs}$ for high-z galaxies. 
  Figure \ref{alma_obs} shows the expected flux for a fiducial ISM mass of 10$^{10}$\msun ~as a function of redshift
for ALMA Bands 3, 4, 6 and 7. These curves can be used to translate our observed fluxes into ISM masses.

\begin{figure*}
\plotone{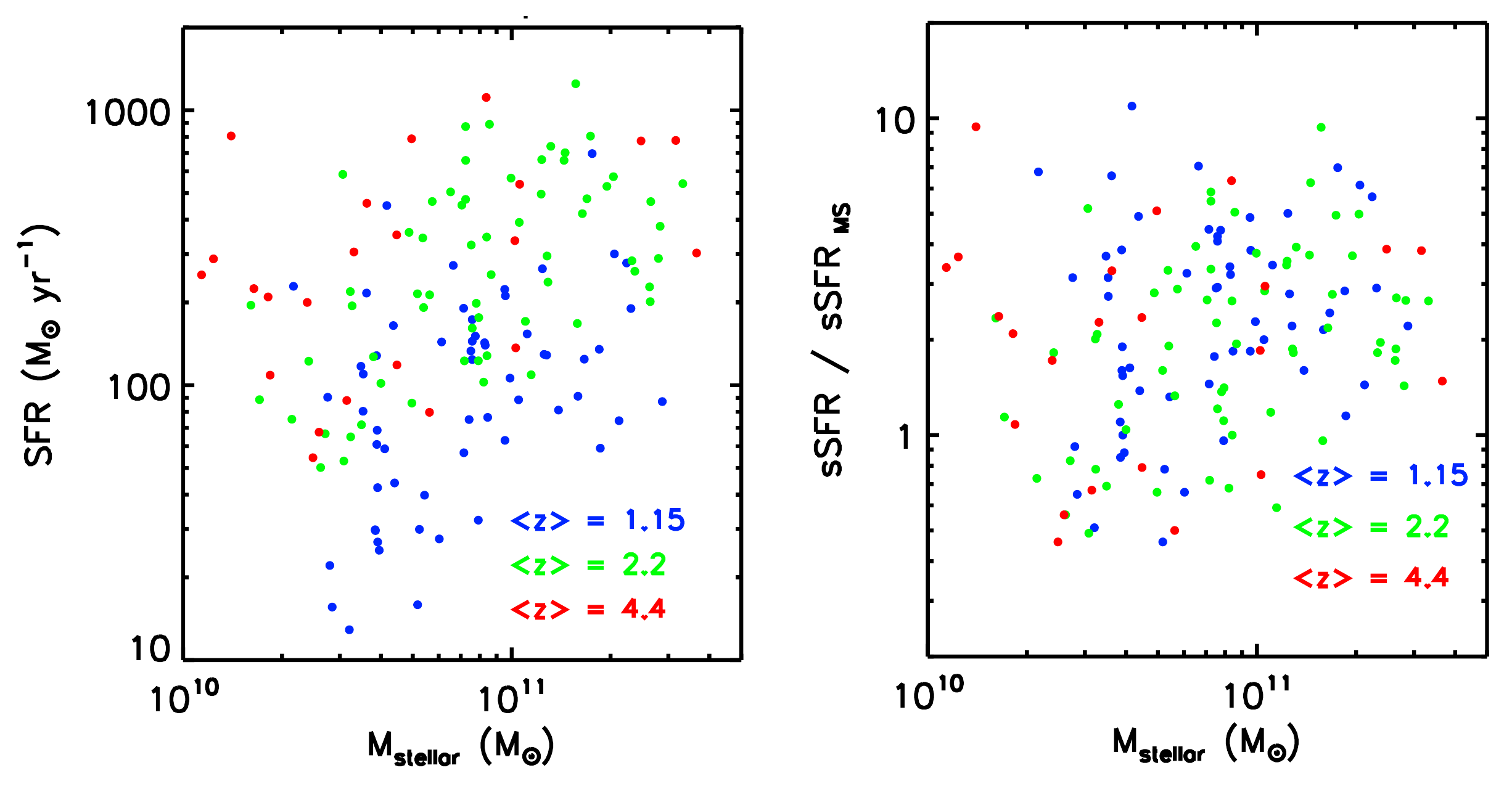}
\caption{Galaxy sample properties -- {\bf Left:} Stellar masses and SFRs for the 145 galaxies in our sample -- 59, 63 and 23 galaxies at each of three redshifts: $< z >$ = 1.15, 2.2 and 4.4. 
SFRs are the sum of the far-infrared from Herschel PACS and SPIRE 
and the unextincted UV of the galaxy. {\bf Right:} The galaxies are shown as a function of their specific star formation rates (sSFR) relative to the sSFR 
of the star-forming Main-Sequence at each galaxy's redshift and stellar mass. The Main-Sequence definition is taken from \cite{lee15} and \cite{sch15} (see text).
At each redshift, the samples probe the variation in ISM molecular gas 
as a function of both stellar mass and SFR from the galaxy MS to $\sim$10 times above the MS. }
\label{sample_fig} 
\end{figure*}

 \begin{figure*}[ht]
\epsscale{1.}  
\plotone{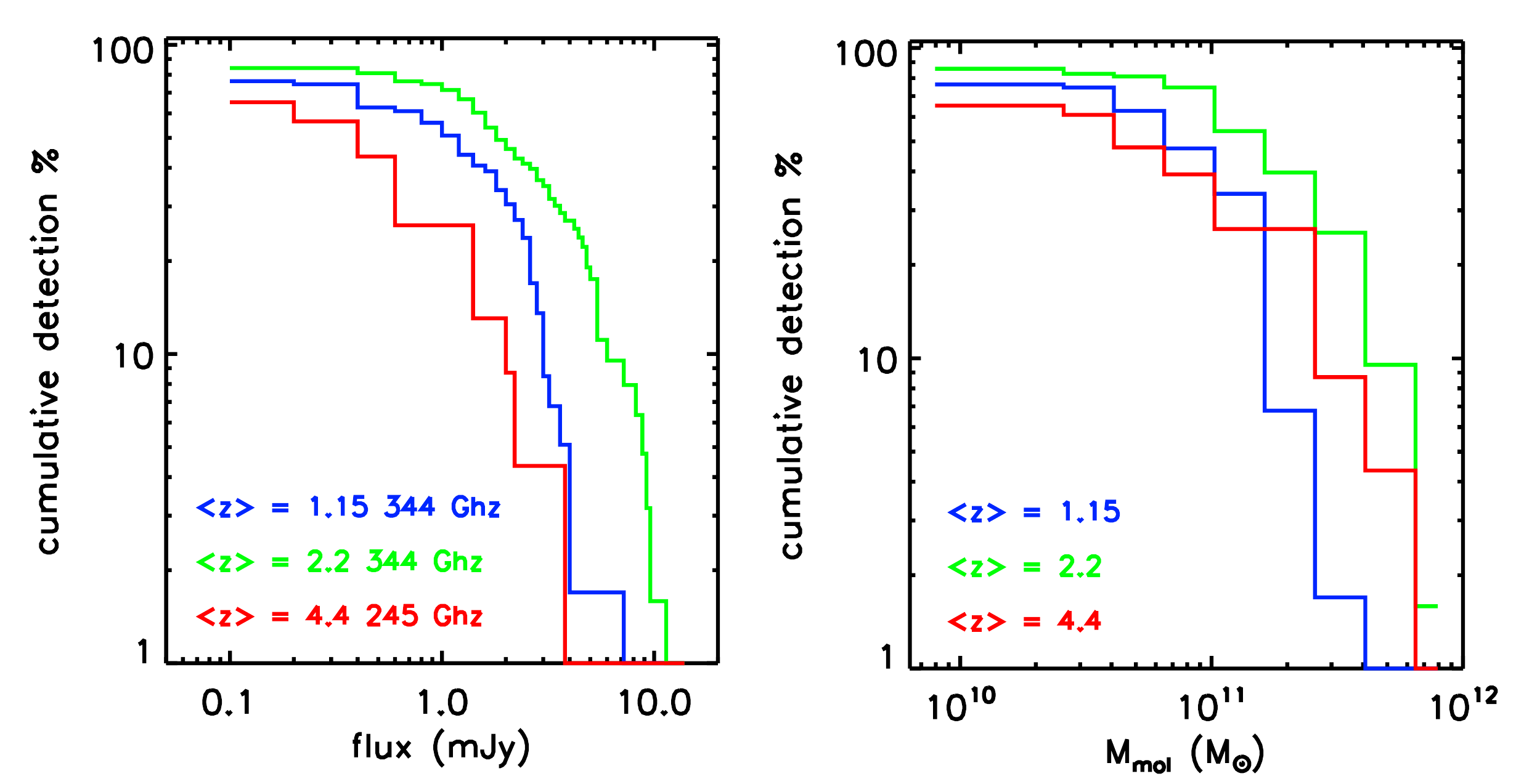}
\caption{{\bf Left:} The cumulative distribution of detection rates as a function of flux
in Band 7 (for z = $\sim1.15$ and 2.2) and Band 6 (for z $\sim$ 4.4). {\bf Right:} The corresponding detection rates for molecular mass using the flux to mass conversion developed in Appendix A and shown in Fig. \ref{alma_obs}.}
\label{detection_rates} 
\end{figure*}

 \begin{figure*}[ht]
\epsscale{1.}  
\plotone{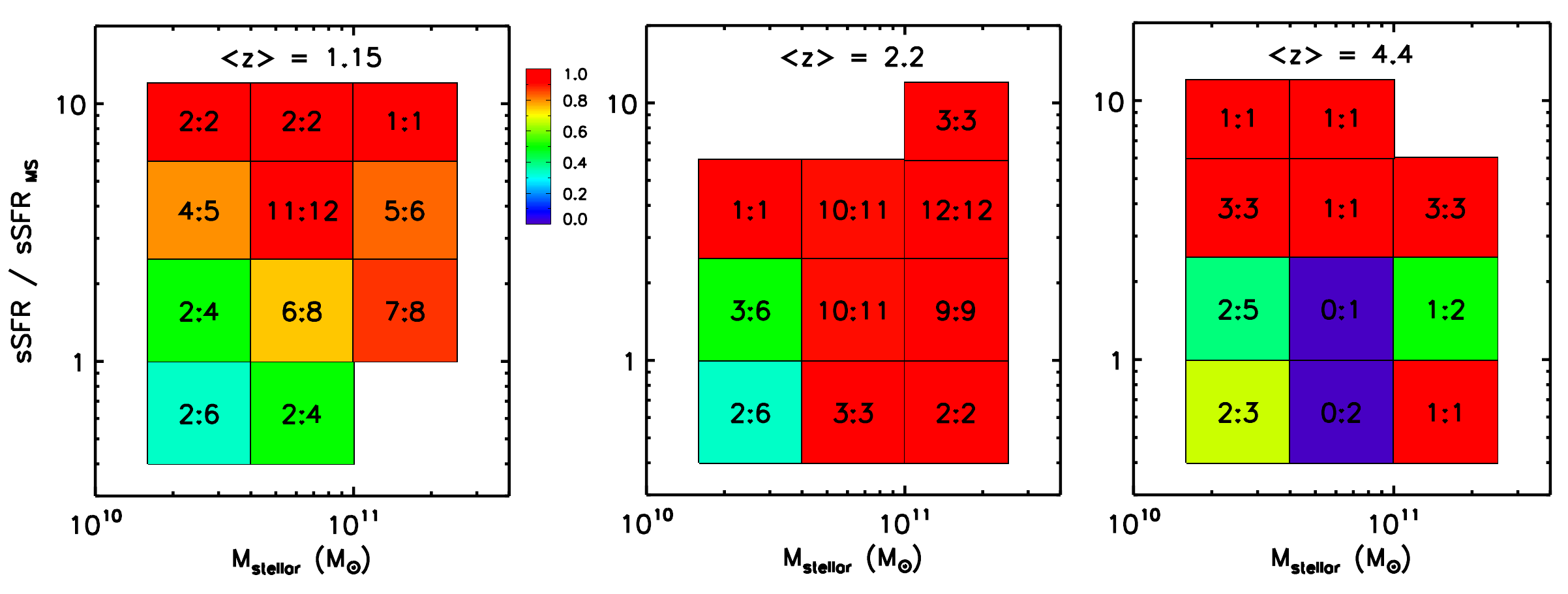}
\caption{Galaxy detection rates as a function of sSFR (relative to the Main Sequence) and M$_{stellar}$ for the three redshift ranges. Each box shows the detection ratio (number of candidates observed : number of detections).  For the MS region (lowest 2 rows), 67\% of the 
observed galaxies are detected; for the higher sSFRs (top 2 rows) the detection rate is 95\%.}
\label{detection_rates1} 
\end{figure*}

\subsection{Dust Temperatures}

Although the submm flux will vary linearly with dust temperature (Equation \ref{fnu}), the range of {\bf mass-weighted} $<T_D>_M$ will be small, 
except in very localized regions. For radiatively heated dust, $T_D$ will vary as the 1/5 - 1/6 power of the 
ambient radiation energy density, implying a 30-fold increase in energy density to double the temperature.
Extensive surveys of nearby galaxies with Herschel find a 
range of $T_D \sim 15 - 30$ K \citep{dun11,dal12,aul13}. Our three calibrations yielding similar values of $\alpha_{850\mu \rm m}$ including normal star-forming and starbursting systems lay a solid foundation for using the Rayleigh-Jeans (RJ) dust emission to probe global ISM masses without introducing a variable 
dust temperature (see Section \ref{temp}). Here we advocate adoption of a single value $<T_D>_M = 25$K (see Section \ref{temp}). 

In fact, it would be wrong to use a variable temperature correction based on fitting to the overall spectral energy 
distribution (SED) since the temperature thus derived is a luminosity-weighted $<T_D>_L$. The difference is easily understood by looking at nearby star-forming Giant Molecular Clouds (GMCs) where spatially resolved far infrared imaging indicates $<T_D>_L\ \sim 40 - 60$K (dominated by the active star-forming centers) whereas 
the overall mass-weighted $<T_D>_M\  \simeq 20$K (dominated by the more extended cloud envelopes. A good illustration of this might be taken from spatially resolved 
far infrared imaging of nearby GMCs. In the Orion, W3, and Auriga Giant Molecular Clouds (GMCs) the far infrared luminosity weighted dust temperature is $\sim50$K and most of that luminosity originates in the few parsec 
regions associated with high mass SF (e.g. M42 and the Kleinmann-Low nebula in the case of Orion). On the other hand, most of the cloud mass is in the extended GMC of 30 - 40 pc diameter and far infrared 
color temperature $\sim15 - 25$K \citep[e.g.][]{mot10,har13,riv15}.  

Within galaxies, there will of course be localized regions  where $T_D$ is significantly elevated -- an extreme
example is the central 100 pc of Arp 220. There, the dust temperatures reach 100 - 200 K \citep{wil14,sco15}; nevertheless, measurements of the whole of Arp 220 are still consistent with the canonical value of $ \alpha_{850\mu \rm m}$ adopted here \citep[see Figure 1 in][]{sco14}.

 \begin{figure*}[ht]
\epsscale{1.2}  
\plotone{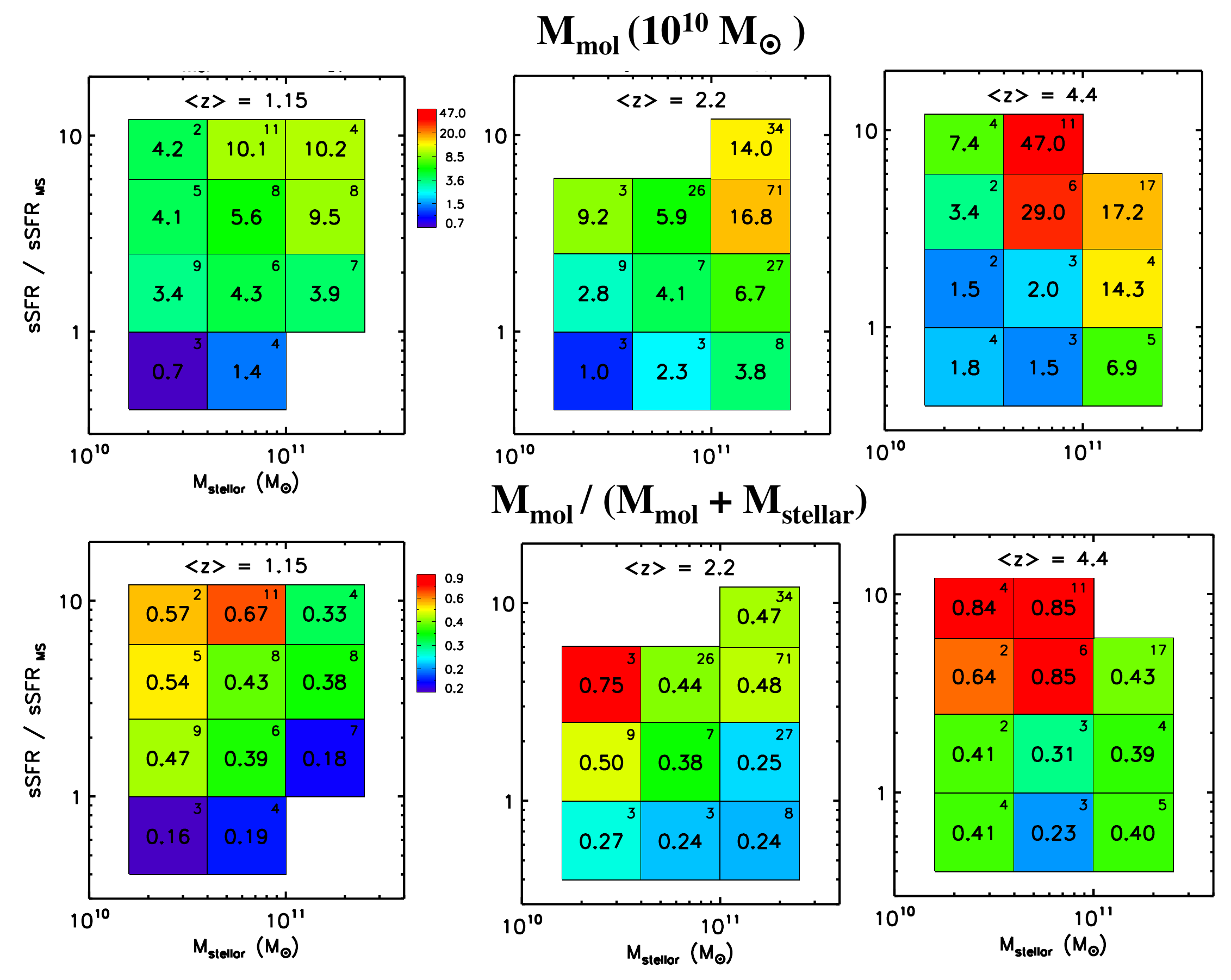}
\caption{Top Row: The gas masses (in units of $10^{10}$\msun) are shown for the stacked images of galaxy sub-samples within each cell of M$_{stellar}$ and sSFR (Table \ref{stacks}).
The number in the upper right of each cell is the statistical significance or signal-to-noise ratio (n$\sigma$) of the estimate. The {\bf ISM masses increase strongly to higher redshift and in galaxies with the highest sSFR}.  Bottom Row: The gas mass fractions (M$_{\rm mol}$ / (M$_{\rm mol}$ + M$_{\rm stellar}$). }
\label{stack_results} 
\end{figure*}

\begin{deluxetable*}{lccccccccccc}[ht]
\footnotesize
\scriptsize
\tablecaption{\bf{Stacked Galaxy Samples}  }

\tablehead{ 
 \colhead{stack} &  \colhead{\#}  & \colhead{$S_{pix}$} & \colhead{$S_{total}$} & \colhead{SNR\tablenotemark{a}}  & \colhead{$< z >$} &  \colhead{$<$ M$_{*}$$ >$} &  
 \colhead{$< $SFR$ >$} & \colhead{$< $sSFR$ >$} &   \colhead{$< $M$_{\rm mol}$$ >$} &   \colhead{$< \tau_{SF}>$} &   \colhead{$M_{\rm mol} /$}\\ 
 \colhead{} & \colhead{galaxies} & \colhead{mJy/beam} & \colhead{mJy}  & \colhead{} & \colhead{}  &  \colhead{ $10^{11}$\msun} &  \colhead{\msun ~yr$^{-1}$}& \colhead{$sSFR_{MS}$} & \colhead{$10^{10}$\msun} & \colhead{Gyr} & \colhead{$(M_{*}+M_{\rm mol})$}}
\startdata 
\\
  \bf{$\rm {\bf <z>}$ = 1.15} \\ \\
        \bf{lowz cell 1}  &   6 &      0.11$\pm$ 0.05 &      0.14$\pm$ 0.05 &      2.96 &  1.06 &      0.34 &   22. &  0.79 &  0.66 &  0.30 &  0.16$\pm$0.054 \\
        \bf{lowz cell 2}  &   4 &      0.13$\pm$ 0.07 &      0.30$\pm$ 0.08 &      3.96 &  1.15 &      0.60 &   26. &  0.70 &  1.41 &  0.54 &  0.19$\pm$0.048 \\
        \bf{lowz cell 4}  &   4 &      0.16$\pm$ 0.08 &      0.73$\pm$ 0.08 &      9.20 &  1.14 &      0.39 &   51. &  1.54 &  3.43 &  0.67 &  0.47$\pm$0.051 \\
        \bf{lowz cell 5}  &   8 &      0.92$\pm$ 0.17 &      0.27$\pm$ 0.05 &      4.99 &  1.15 &      0.69 &   67. &  1.72 &  4.34 &  0.65 &  0.39$\pm$0.070 \\
        \bf{lowz cell 6}  &   8 &      0.81$\pm$ 0.12 &      0.40$\pm$ 0.04 &      8.98 &  1.21 &      1.75 &   91. &  1.86 &  3.93 &  0.43 &  0.18$\pm$0.027 \\
        \bf{lowz cell 7}  &   5 &      0.86$\pm$ 0.16 &      0.39$\pm$ 0.06 &      6.47 &  1.12 &      0.35 &  105. &  3.37 &  4.07 &  0.39 &  0.53$\pm$0.101 \\
        \bf{lowz cell 8}  &  12 &      1.17$\pm$ 0.14 &      0.43$\pm$ 0.04 &     11.26 &  1.18 &      0.75 &  154. &  3.74 &  5.61 &  0.37 &  0.43$\pm$0.052 \\
        \bf{lowz cell 9}  &   6 &      1.94$\pm$ 0.23 &      0.42$\pm$ 0.06 &      7.47 &  1.23 &      1.54 &  178. &  3.60 &  9.46 &  0.53 &  0.38$\pm$0.046 \\
       \bf{lowz cell 10}  &   2 &      0.85$\pm$ 0.34 &      0.79$\pm$ 0.09 &      8.96 &  1.23 &      0.31 &  221. &  6.65 &  4.17 &  0.19 &  0.57$\pm$0.230 \\
       \bf{lowz cell 11}  &   2 &      2.07$\pm$ 0.20 &      0.73$\pm$ 0.09 &      8.19 &  1.25 &      0.50 &  390. &  9.61 & 10.06 &  0.26 &  0.67$\pm$0.063 \\
       \bf{lowz cell 12}  &   1 &      2.11$\pm$ 0.50 &      1.46$\pm$ 0.15 &      9.64 &  1.20 &      2.06 &  300. &  6.15 & 10.19 &  0.34 &  0.33$\pm$0.079 \\
 
\\  
      \bf{$\rm {\bf <z>}$ = 2.2} \\ \\
        \bf{midz cell 1}  &   6 &      0.15$\pm$ 0.05 &      0.19$\pm$ 0.06 &      3.07 &  2.20 &      0.28 &   64. &  0.67 &  1.05 &  0.16 &  0.27$\pm$0.088 \\
        \bf{midz cell 2}  &   3 &      0.27$\pm$ 0.08 &      0.26$\pm$ 0.05 &      5.41 &  2.73 &      0.71 &  115. &  0.71 &  2.26 &  0.20 &  0.24$\pm$0.074 \\
        \bf{midz cell 3}  &   2 &      0.39$\pm$ 0.05 &      0.31$\pm$ 0.05 &      5.67 &  2.66 &      1.21 &  117. &  0.64 &  3.78 &  0.32 &  0.24$\pm$0.031 \\
        \bf{midz cell 4}  &   6 &      0.38$\pm$ 0.02 &      0.40$\pm$ 0.04 &      9.30 &  2.20 &      0.28 &  173. &  1.82 &  2.80 &  0.16 &  0.50$\pm$0.054 \\
        \bf{midz cell 5}  &  11 &      0.75$\pm$ 0.11 &      0.69$\pm$ 0.05 &     15.24 &  2.24 &      0.68 &  191. &  1.48 &  4.09 &  0.21 &  0.37$\pm$0.056 \\
        \bf{midz cell 6}  &   9 &      1.05$\pm$ 0.06 &      1.22$\pm$ 0.05 &     26.79 &  2.25 &      2.01 &  266. &  1.75 &  6.70 &  0.25 &  0.25$\pm$0.009 \\
        \bf{midz cell 7}  &   1 &      1.66$\pm$ 0.55 &      0.90$\pm$ 0.15 &      6.01 &  2.34 &      0.31 &  585. &  5.19 &  9.17 &  0.16 &  0.75$\pm$0.246 \\
        \bf{midz cell 8}  &  11 &      0.87$\pm$ 0.03 &      0.94$\pm$ 0.04 &     26.12 &  2.43 &      0.74 &  608. &  4.02 &  5.90 &  0.10 &  0.44$\pm$0.017 \\
        \bf{midz cell 9}  &  12 &      2.56$\pm$ 0.04 &      3.06$\pm$ 0.04 &     70.57 &  2.30 &      1.82 &  559. &  3.56 & 16.83 &  0.30 &  0.48$\pm$0.007 \\
       \bf{midz cell 12}  &   3 &      1.79$\pm$ 0.14 &      2.60$\pm$ 0.08 &     34.13 &  1.94 &      1.60 &  885. &  7.56 & 13.99 &  0.16 &  0.47$\pm$0.014 \\

\\   
  \bf{$\rm {\bf <z>}$ = 4.4} \\ \\

       \bf{highz cell 1}  &   3 &      0.11$\pm$ 0.05 &      0.15$\pm$ 0.04 &      3.99 &  4.71 &      0.27 &   70. &  0.57 &  1.84 &  0.26 &  0.40$\pm$0.100 \\
       \bf{highz cell 2}  &   2 &      0.08$\pm$ 0.04 &      0.12$\pm$ 0.04 &      2.75 &  4.23 &      0.51 &   97. &  0.63 &  1.50 &  0.15 &  0.23$\pm$0.082 \\
       \bf{highz cell 3}  &   1 &      0.55$\pm$ 0.11 &      0.20$\pm$ 0.06 &      3.54 &  4.04 &      1.03 &  137. &  0.75 &  6.90 &  0.50 &  0.40$\pm$0.078 \\
       \bf{highz cell 4}  &   5 &      0.12$\pm$ 0.06 &      0.10$\pm$ 0.03 &      3.72 &  4.63 &      0.22 &  211. &  1.90 &  1.52 &  0.07 &  0.40$\pm$0.191 \\
       \bf{highz cell 5}  &   1 &      0.13$\pm$ 0.06 &      0.16$\pm$ 0.06 &      2.59 &  5.59 &      0.45 &  352. &  2.35 &  2.00 &  0.06 &  0.31$\pm$0.119 \\
       \bf{highz cell 6}  &   2 &      1.17$\pm$ 0.26 &      0.67$\pm$ 0.04 &     15.59 &  4.52 &      2.21 &  321. &  1.68 & 14.33 &  0.45 &  0.39$\pm$0.088 \\
       \bf{highz cell 7}  &   3 &      0.27$\pm$ 0.11 &      0.15$\pm$ 0.04 &      4.13 &  4.25 &      0.19 &  328. &  3.44 &  3.40 &  0.10 &  0.64$\pm$0.262 \\
       \bf{highz cell 8}  &   1 &      2.24$\pm$ 0.40 &      1.33$\pm$ 0.06 &     23.27 &  3.54 &      0.50 &  788. &  5.10 & 28.95 &  0.37 &  0.85$\pm$0.152 \\
       \bf{highz cell 9}  &   3 &      1.37$\pm$ 0.08 &      0.94$\pm$ 0.04 &     23.98 &  4.03 &      2.24 &  696. &  3.54 & 17.16 &  0.25 &  0.43$\pm$0.025 \\
      \bf{highz cell 10}  &   1 &      0.60$\pm$ 0.14 &      0.24$\pm$ 0.07 &      3.60 &  4.18 &      0.14 &  807. &  9.40 &  7.45 &  0.09 &  0.84$\pm$0.201 \\
      \bf{highz cell 11}  &   1 &      3.84$\pm$ 0.35 &      2.49$\pm$ 0.06 &     39.14 &  4.64 &      0.84 & 1114. &  6.35 & 47.03 &  0.42 &  0.85$\pm$0.077 \\
       
  \\ 
\enddata\label{stacks}
\tablecomments{
$\rm \tau_{SF} = M_{\rm mol} / <SFR>$ where $\rm <SFR>$ is the mean SFR.}
\tablenotetext{a}{SNR is the higher of the SNR$_{\rm tot}$ and SNR$_{\rm peak}$ (see text). }
\end{deluxetable*}

\section{Galaxy Samples for ALMA}\label{sample}

Our sample of 145 galaxies is taken from the COSMOS 2 deg$^2$ survey \citep{sco_ove}. This survey field has excellent photometric redshifts \citep{ilb13,lai15} derived from deep 34 band (UV-Mid IR) photometry. The galaxies were selected to sample stellar masses $M_{stellar}$ in the range $0.2- 4\times10^{11}$\msun\ and the range of SFRs at each stellar mass. This
is not a representative sampling of the galaxy population but rather meant to cover the range of galaxy properties. Here, 55\% of the galaxies are within a factor 2.5 of the SFR on the MS, whereas for the overall population of SF galaxies at z $\sim$ 2, there is a much larger fraction. Forty-eight have spectroscopic redshifts and 120 have at least a single band detection in the infrared 
with Spitzer MIPS-24$\mu$m or Herschel PACS and SPIRE; 65 had two or more band detections with Herschel PACS/SPIRE.  

The photometric redshifts and stellar masses of the galaxies are from \cite{mcc12,ilb13,lai15}. 
Preference was given to the most recent photometric redshift catalog \citep{lai15} which makes use of deep Spitzer SPASH IRAC imaging \citep{ste14} and the latest release of COSMOS UltraVista. The SFRs assume a Chabrier stellar initial mass function (IMF); they are derived from the rest frame UV continuum and infrared using Herschel PACS and SPIRE data as detailed in \cite{ilb13} and in \cite{lee15} . For sources with detections in at least at least two of the five available Herschel bands, $L_{IR}$  is estimated by fitting far-infrared photometry to a coupled, modified greybody plus mid-infrared power law, as in \cite{cas12}. The mid-infrared power-law slope and dust emissivity are fixed at $\alpha$ = 2.0 and $\beta$ = 1.5, respectively. 

The original sample of galaxies observed with ALMA had 180 objects. However, subsequent to the ALMA observations, 
new photometric redshifts \cite{lai15} and analysis of the Hershcel PACs and SPIRE measurements in COSMOS \citep{lee15} became available. We have made use of those new ancillary data to refine the sample, including only those objects with most reliable redshifts, stellar masses and SFRs (as judged from the photometric redshift fitting uncertainties). We also required that the derived stellar masses agree within a factor 2 between the two most recent photometric redshift catalogs \citep{ilb13,lai15}.  The individual objects are tabulated in Appendix \ref{source_app}. There, the adopted redshifts, stellar masses and SFRs for each of the individual galaxies are tabulated in Tables \ref{lowz} - \ref{highz}. 

For each galaxy, we also list the specific star formation rate relative to that of the MS at the same redshift and stellar mass (sSFR$_{MS}$ / sSFR$_{MS}$). In recent years, there have been numerous works specifying the MS evolution \citep{noe07,rod11,bet12,spe14,lee15,sch15}. The last two works have similar specification of the MS as a function of stellar mass at low redshift. Here, we use \cite{lee15} with no evolution of the MS beyond z = 2.5 (i.e. MS(z $>$ 2.5) = MS(z = 2.5). The \cite{lee15} MS was adopted here since the infrared-based SFRs were also taken from \cite{lee15}; thus the SFRs will have the same calibration.  The three sub-samples with 59, 63 and 23 galaxies at $\rm < z > \sim$ 1.15, 2.2 and 4.4, respectively, probe SFRs from the MS up to $10\times sSFR_{MS}$ with $\rm M_{stellar} = 0.2 - 4\times10^{11} \msun$ (Figure \ref{sample_fig} and Table \ref{stacks}).

\section{Observations and Flux Measurements}\label{obs}

The ALMA Cycle 2 observations (\#2013.1.00034.S) were obtained in 2014-2015. The z = 1.15 \& 2.2 samples were observed in 
Band 7 (345 GHz), the z = 4.4 sample in Band 6 (240 GHz). On-source integration times were $\sim2$ minutes per galaxy and average rms sensitivities were 0.152 (Band 7) and 0.065 mJy beam$^{-1}$ (Band 6). Synthesized beam sizes were $\simeq0.6 - 1$\arcsec. Data were calibrated and imaged with natural weighting using CASA.

The detection rates are summarized in Figure \ref{detection_rates} as a function of flux (Left Panel) and ISM mass (Right panel) and in Figure \ref{detection_rates1} as a function of M$_{stellar}$ and sSFR. 
The detection rates for individual galaxies are $\sim$70, 85 and 50\% at z = 1.15, 2.2 and 4.4 respectively. All flux measures are restricted to within 1.5\arcsec ~on the galaxy position. 
%First, we used the SourceExtractor program \citep{ber96} to search for emission peaks within the central 3\arcsec aperture. The detection threshold (typically $\sim2.8 - 3\sigma$)
%was adjusted for each image so that the probability of spurious detection would be less than unity in the entire sample of 180 images. 
For detections, we searched for significant peaks or aperture-integrated flux within the central 3\arcsec ~surrounding each program source. We required a 2$\sigma$ detection in S$\rm_{tot}$ or 3.6$\sigma$ in S$\rm_{peak}$, in order that the detection be classified as real. These limits ensure that there would be less than one spurious detection in the145 objects.
Noise estimates for the integrated flux measures were derived from the dispersion in the aperture-integrated fluxes 
for 100 equivalent apertures, displaced off-source in the same image.

 \section{Stacked Samples}
 
Here we focus on results derived from stacking the images of subsamples of galaxies in cells of M$_{stellar}$ and sSFR (Figure \ref{detection_rates1}). 
%\footnote{The measurements for the individual galaxies will be presented and analyzed in a later paper. That paper will include a full discussion of uncertainties in the galaxy properties and the photometric redshifts in addition to new spectroscopic redshifts from Keck MOSFIRE.} 
The galaxy images of all galaxies 
in each cell were both median- and average-stacked. Given the small numbers of galaxies in many of the sub-samples (see Figure \ref{detection_rates1}), we  used the 
average stack rather than the median; for such small samples the median can have higher dispersion. 
Flux and mass measurements for the stacked subsamples of galaxies are given in Table \ref{stacks} along with the mean sSFR and M$_{\rm stellar}$ of each cell.  

\subsection{Gas Masses}
 
 Figure \ref{stack_results} shows derived mean gas masses and gas mass fractions of each cell for the three redshifts.  The values for $M\rm_{\rm mol}$ and the
 gas mass fraction (M$\rm_{\rm mol}$ / (M$\rm_{\rm mol}$ + M$\rm_{stellar}$)) are given by the large numbers in each cell and the statistical significance is given by the smaller 
 number in the upper right of each cell.
 
 Figure \ref{stack_results}-Top shows a large increase in the ISM masses from z = 1.15 to z = 2.2 for galaxies with stellar mass $\geq10^{11}$ \msun ~and for galaxies with sSFR/sSFR$_{MS} \geq 4$ (i.e. galaxies in the upper right of the diagrams).  For lower mass galaxies and galaxies at or below the MS, less evolutionary change is seen, although the MS is itself evolving upwards in sSFR. From z = 2.2 to 4.4, there is milder evolution since approximately equal numbers of cells have higher and lower M$_{\rm mol}$ at z = 4.4 compared with z = 2.2 and the differences don't appear strongly correlated with sSFR and M$\rm_{stellar}$. 
 
 \subsection{Gas Mass Fractions}
 
 The gas mass fractions shown in Figure \ref{stack_results}-Bottom range from $\sim0.2 - 0.5$ on the MS (bottom two rows of cells) up to 0.5 - 0.8 for the highest sSFR cells. 

For perspective, we note that the Milky Way galaxy has a stellar mass $\simeq6\times10^{10}$\msun \citep{mcm11}, M$_{\rm ISM} \sim3 - 6\times10^9$\msun \ (approximately equally contributed by HI and H$_2$) and SFR $\sim1 - 2$ \msun ~yr$^{-1}$. Thus for the Galaxy, sSFR $\simeq 0.015 - 0.030$ Gyr$^{-1}$ and gas mass fraction is $\sim$0.055. Lastly, to place the Milky Way in context at low z, the main sequence parameters given by \cite{bet12} and \cite{lee15} yield SFR = 4.2 and 3.8 \msun ~yr$^{-1}$ and sSFR = 0.07 and 0.063 Gyr$^{-1}$ for the Milky Way's stellar mass; thus, the Galaxy has a SFR $\sim$2 times below the z = 0 MS but is still classified as a MS galaxy. 

Compared to the Galaxy, the MS galaxies at z $= 1 - 6$  
therefore have $\sim5 - 10$ times higher gas mass fractions for the same stellar mass and $\sim100$ times higher gas masses in the highest stellar mass galaxies. At low redshift, such massive 
galaxies ($M_{stellar} = 4\times10^{11}$ \msun) would have largely evolved to become passive (non-star forming) red galaxies with much lower ISM masses.  

The trends in gas masses, SFRs and gas mass fractions can be represented adequately by quite simple analytic functions. Using the IDL 
implementation of the Levenberg-Marquardt algorithm for non-linear least squares fitting (LMFIT) of the  
data shown in Figure \ref{stack_results} and Table \ref{stacks}, we obtain:
 \begin{eqnarray}
  {\rm M_{\rm mol} \over {M_{\rm mol} + M_{stellar}}} &=& ~(0.30\pm 0.02)~\left({\rm M_{\rm stellar} \over 10^{11}\msun}\right)^{-0.02 \pm 0.02} ~    \nonumber \\
 && \times  \left({\rm 1+z \over 3}\right)^{0.44 \pm 0.05}  \rm \left({\rm sSFR \over sSFR_{MS}}\right)^{0.32 \pm 0.02} .  \label{gas-fraction_law} 
 \end{eqnarray} 
 
 \noindent ; the parameter uncertainties in Equation \ref{gas-fraction_law} are those obtained from the Levenberg-Marquardt algorithm. We also attempted fitting the gas mass fractions with a sSFR / sSFR$_{MS}$ term; this did not improve the fit and we therefore kept the simpler, un-normalized SFR term. 

The gas mass fractions derived here are quite consistent with values derived in a number of other studies from observations of 
CO (mostly 2-1 and 3-2 line). \cite{tac10} obtained a range of 0.2 - 0.5 at z $\sim 1.1$ and 0.3 -0.8 at z $\sim 2.3$. \cite{dad10} estimated a gas mass fraction $\sim 0.6$ for 6 galaxies at z = 1.5 
and \cite{mag12a} measured 0.36 in a Lyman break galaxy at z = 3.2.  Later, more extensive studies were done by: \cite{tac13} with 52 galaxies and mean gas fractions of 0.33 and 0.47 at 
z $\sim 1.2$ and 2.2, respectively;  \cite{san14}

 Several cells in the sSFR-M$_{stellar}$ plane have gas mass fractions 50 - 80\%, implying gas masses 1 -- 4 times the stellar masses. These galaxies have the highest sSFRs at z = 2.2 and 4.4. Clearly, {\bf such 
galaxies can not be made from the merging of two main sequence galaxies} having gas mass fractions $\sim 40$\%, since in a merger the resultant gas mass fraction would remain constant 
or even decrease (if there is significant conversion of gas to stars in a starburst). 

These gas-dominated galaxies with very high sSFR galaxies must therefore indicate a different aspect of galaxy evolution -- perhaps either {\bf nascent galaxies} -- having M$_{\rm mol} \rm > M_{stellar}$ (yet clearly having prior star formation given their large stellar masses and the presence of metal enriched ISM), or galaxies in environments yielding very high IGM accretion rates. These galaxies share the gas-rich properties of the submillimeter galaxies, yet the ones seen here were selected first in the optical-NIR without pre-selection for dust emission. The gas masses of these systems reach $4\times10^{11}$ \msun~ -- they are very likely the progenitors of the present epoch massive elliptical galaxies \citep{tof14}. 

\subsection{Star Formation Law at High Redshift}
 
 \begin{figure*}[ht]
\epsscale{1.}  
\plotone{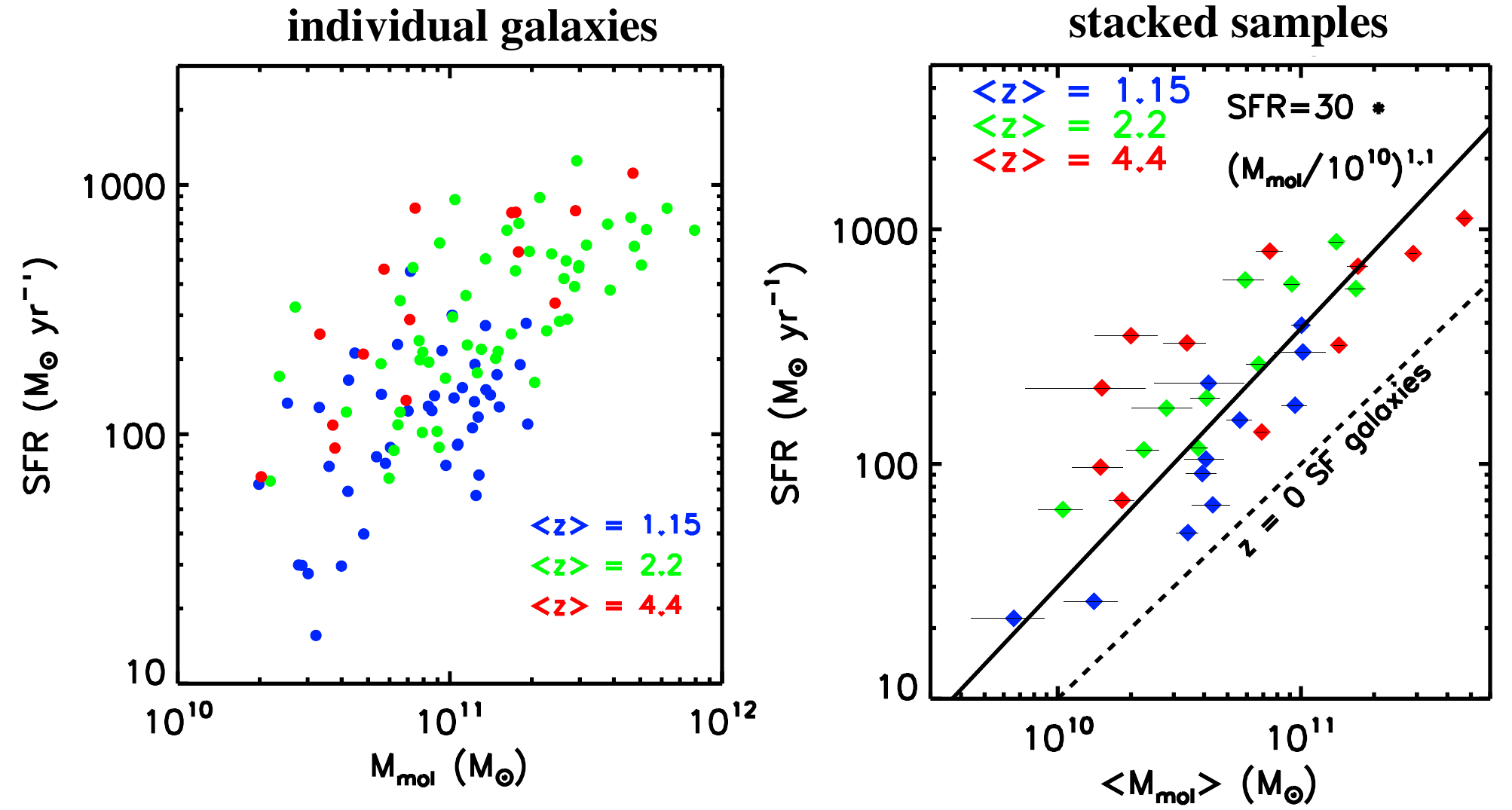}
\caption{{\bf Left:} The SFRs as a function of $M_{mol}$ are shown for the individual galaxies and on the {\bf Right} for the stacked galaxy samples. The best fit SF law, given by Eq. \ref{sfr_law}
and evaluated at z = 2, is shown in the right panel.}
\label{sfr_indiv} 
\end{figure*}

 \begin{figure}[ht]
\epsscale{1.}  
\plotone{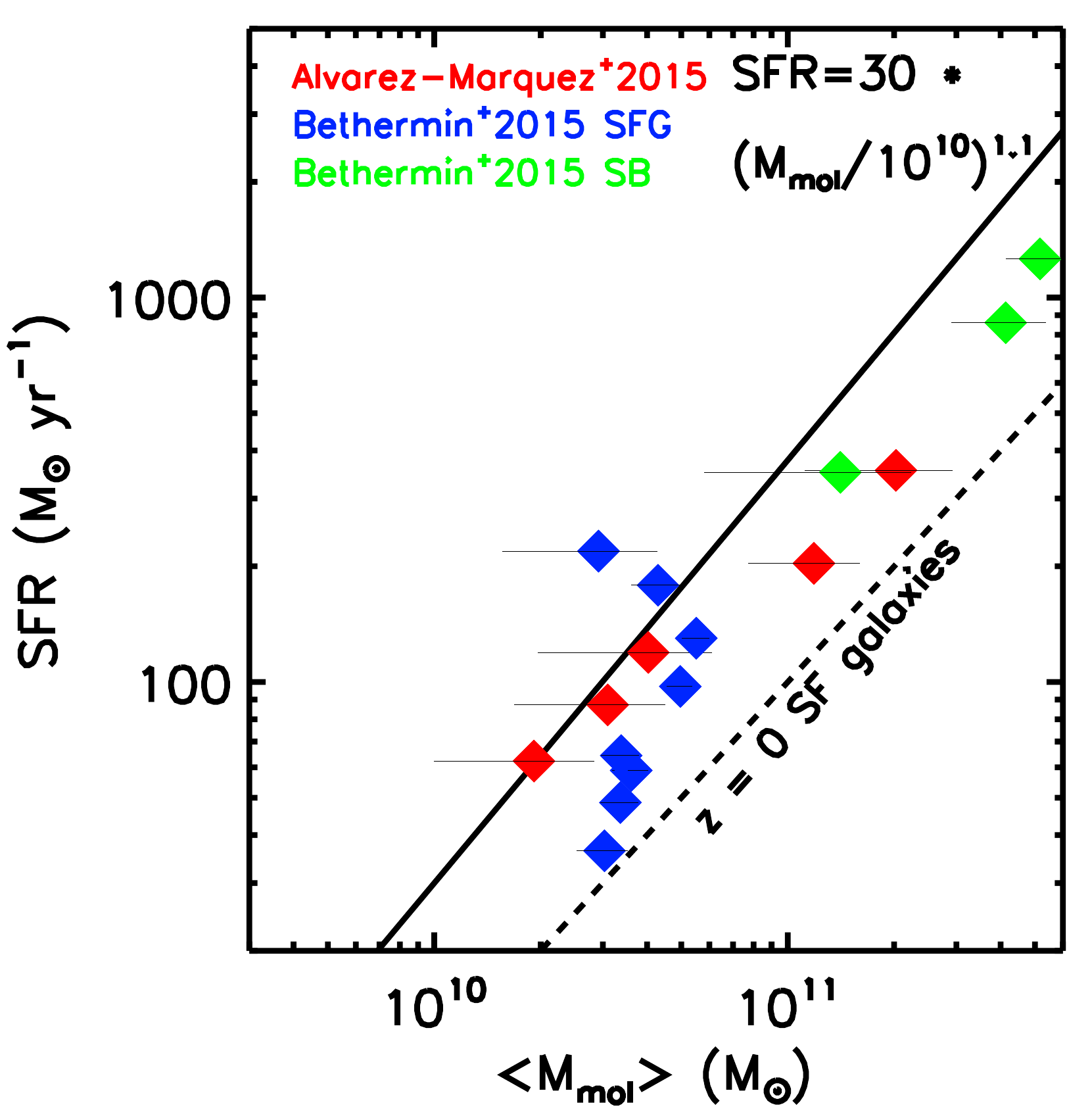}
\caption{The best fit SF law given by Eq. \ref{sfr_law} (evaluated for z = 2) is compared with the stacking results from \cite{alv15} and \cite{bet15}, indicating 
reasonably good agreement considering the very different approaches. \cite{alv15} and \cite{bet15} did stacking of very large samples of galaxies using 
Herschel SPIRE 350 and 500$\mu$m and Aztec 1.1 mm imaging. Here we make use of only their 1.1 mm stack fluxes in order to stay on the RJ tail and convert those fluxes 
to an implied ISM molecular mass using the identical procedure developed here for translating submm flux density into mass.}
\label{sfr_stacks} 
\end{figure}

Using measurements from the stacking in cells of sSFR and M$_{stellar}$ (Table \ref{stacks}), we obtained a least-squares fit for the 
SFR dependence on gas mass, redshift and elevation above the main sequence:

 \begin{eqnarray}
 \rm SFR  &=& (35\pm 16) ~  \left({\rm M_{\rm mol} \over 10^{10}~\msun}\right)^{0.89 \pm 0.12} ~\times    \nonumber \\
 && \left({\rm1+z\over3}\right)^{0.95\pm0.28}  \left({\rm sSFR\over sSFR_{MS}}\right)^{0.23\pm0.15} \msun \rm yr^{-1} . \label{sfr_law}
  \end{eqnarray}
  
\noindent ; the parameter uncertainties were obtained from the Levenberg-Marquardt procedure. 

Equation \ref{sfr_law} indicates a high redshift SF law with an approximately linear dependence on gas mass and an increasing SF efficiency (SFR per unit gas mass) at higher redshift. The dependence on sSFR relative to that on the MS is relatively weak (0.23 power; see also Section \ref{mass}). No significant dependence (less than 1 $\sigma$) on stellar mass was found so we omitted the stellar mass term from the fitting for Equation \ref{sfr_law}. 
Inversion of Equation \ref{sfr_law} yields an expression for the gas mass in terms of the observed SFRs. 

For low redshift galaxies, \cite{ler13} derive a very similar power law index (0.95 - 1) for the SF as a function of molecular gas mass but with a lower SF rate per unit ISM mass;
this result was also seen in earlier surveys of nearby galaxies in CO \citep[see][]{you91}.

In Figure \ref{sfr_stacks}, we compare the SF law given in Equation \ref{sfr_law} with the stacking results of \cite{bet15} and \cite{alv15}. Their 1.1 mm stacked fluxes were translated 
into gas masses using the procedure developed here in Appendix \ref{dust_app}. \cite{bet15} had samples of galaxies at z $>$ 1 both on the MS and at higher SFRs; 
the \cite{alv15} sample has Lyman Break galaxies at z $\sim$ 3, presumably mostly MS galaxies. The 1.1 mm measurements on which the mass estimates in Figure \ref{sfr_stacks} are based used very large beams and therefore had many sources within the beam; this confusion was removed statistically \citep[see][]{bet15,alv15}. (We do not use their SPIRE 500$\mu$m stacked fluxes since for most of these 
redshifts $\lambda_{rest}$ will be less than 250$\mu$m and thus not safely on the RJ tail.) Figure \ref{sfr_stacks} indicates reasonable agreement between the ALMA results presented here and the 1.1 mm stacking results; both show an approximately linear dependence of the SFRs on the estimated gas masses.  

\subsection{Gas Depletion Times}

\begin{figure*}[ht]
\epsscale{1.}  
\plotone{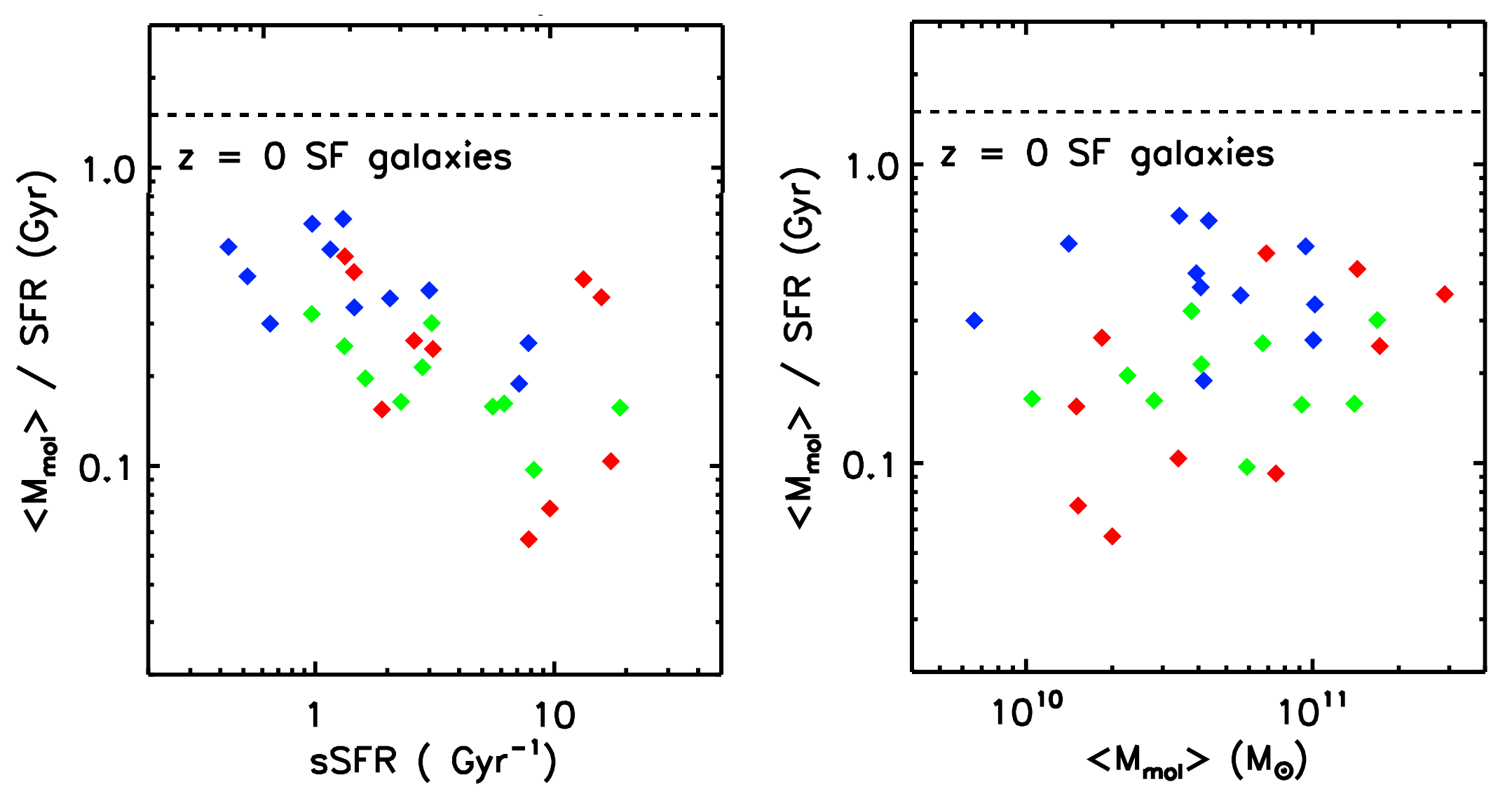}
\caption{The gas depletion times ($\tau_{dep} = M_{mol} / SFR$) are shown as a function of sSFR  and M$_{\rm mol}$. }
\label{depletion_time} 
\end{figure*}

The existence of a {\bf single `linear' relation between the available gas mass and the SFR} for {\bf all} our galaxies (independent of redshift at z $>$ 1, both on and above the MS) is a result with fundamental implications.
The characteristic ISM depletion time $\rm \tau = M_{\rm mol} / SFR \simeq 2 - 7\times10^8$ yrs (Figure \ref{depletion_time}) is approximately constant for galaxies both on and above the MS. 
These are broadly consistent which previously determined typical values. 
\cite{tac13} found a mean gas depletion time of $\sim 7\times10^8$ yrs for a sample of 53 galaxies with CO at z = 1 to 2.5 and \cite{san14} found $\sim 1 - 3\times10^8$ yrs for a large sample using 
dust continuum measurements from Herschel (see their Figure 7). 

The gas depletion time determined here does show evolution with redshift, having shorter timescales at z = 2.2 and shorter still at z = 4.4 compared to z = 1.1. 
This timescale is short compared with the $\sim2$ Gyr time differences between z = 4.4 and 2.2 and z = 2.2 to 1.15, implying that there must be substantial 
accretion of fresh gas to replace that being absorbed into stars.

For the nearby galaxies, the gas depletion times are $\sim1.5$ Gyr \citep{you91,you95,ler13}. The shorter depletion times at high redshift and 
the universality of these short timescales imply that star formation in the early universe is driven by very different processes than those in present day galaxies having low star formation efficiencies.
This different {\bf star formation mode, dominant in high redshift gas-rich galaxies, is quite plausibly the same dynamically driven SF occurring in low-z galactic spiral arms, bars and merging systems}. However, at high redshift the higher SF efficiencies occur throughout the SF galaxy population. At high redshift, dispersive gas motions (as opposed to ordered rotation) and/or galaxy interactions will lead to compression in the highly dissipative ISM, enhancing the SFRs per unit gas mass. Clearly, such motions will be damped on a galaxy crossing timescale 
($\sim2\times10^8$ yr), but this is also the timescale that is implied for replenishment of the high-z gas masses in order to maintain the observed SFRs. The accreted gas from the intergalactic medium or galaxy merging should have typical infall/free-fall velocities of a few$\times100$ \kms; i.e. sufficient to maintain the dispersive ISM velocities which then can drive the higher SFRs, elevated relative to z = 0 quiescent galaxies.

 \section{On versus above the MS}\label{mass}

In Table \ref{summary_table}, we include average properties and gas masses for samples of galaxies on and above the MS for the three redshift ranges. 
At all three redshifts, the rate of SF per unit gas mass is similar for the MS galaxies and for galaxies with sSFR greater than 2.5 times above the MS (last column in Table \ref{summary_table}).
In the samples above the MS, the sSFRs are typically 3 - 4 times above the MS samples, yet the gas depletion times are less than a factor 2 different. And for the 
complete sample at all redshifts (last three rows in Table \ref{summary_table}) the depletion times differ by only $\sim$\ 20\%. 

Comparing the gas masses and depletion times on and above the MS in Table \ref{summary_table}, it is clear that the higher SFRs for galaxies with 
elevated sSFRs are mostly due to those galaxies having much higher gas masses, rather than an increased efficiency, or rate of converting gas to 
stars. This conclusion is substantiated by fitting function results (Equation \ref{sfr_law}), giving a power-law dependence of 0.89 
on the gas mass and only 0.23 on the sSFR relative to the MS. Thus, the so-called starburst population is largely a population of very 
gas-rich galaxies rather than galaxies converting gas to stars more efficiently. In fact, \cite{mag12} similarly find that the vertical spread of the MS  
band is due to variations in the gas mass fraction rather than variations in the SF efficiency (see their Section 6.3). On the other hand, \cite{sil15} reach a different conclusion with 
CO (2-1) observations for seven z = 1.6 galaxies having sSFR approximately 4 times above the MS. They attribute their lower CO to far infrared luminosity ratios 
to a higher star formation efficiency relative to galaxies on the MS. However, the offset shown in their Figure 3a is really very small -- less than 
a 30\% departure from the CO/FIR ratios occurring on the MS.

\begin{deluxetable*}{ccccccccc}
%\tabletypesize{0.2}
%\normaltsize
%\rotate
\tablecaption{\bf{Averages for Samples on and above MS}}
\tablewidth{0pt}

\tablehead{ 
Sample &  \colhead{\#} &  \colhead{$ < z > $} & \colhead{$< M_{*} >$} & \colhead{$< SFR >$} &\colhead{$< sSFR >$} &   \colhead{$< M_{\rm mol} > $} &   \colhead{$< f_{\rm mol} >$} &   \colhead{$< M_{\rm mol} / SFR >$}   \\
&  \colhead{gal.}  &  \colhead{}  &  \colhead{ $10^{11}$\msun} & \colhead{\msun yr$^{-1}$} & \colhead{$/ sSFR_{MS} $} & \colhead{$10^{10}$\msun}  & \colhead{} & \colhead{Gyr} }
\startdata 

\boldmath$< z > = 1.1$\unboldmath \\

MS &       19 &   1.16 &   0.92 &  64.20 &   1.48 &     6.3$\pm$0.8 &     0.42$\pm$0.04 &     1.09$\pm$0.13 \\ 

above MS &       25 &   1.20 &   0.81 & 169.15 &   4.49 &    10.6$\pm$0.9 &     0.55$\pm$0.04 &     0.65$\pm$0.07 \\ 

all &       44 &   1.19 &   0.88 & 123.69 &   3.15 &     9.0$\pm$0.7 &     0.50$\pm$0.03 &     0.84$\pm$0.07 \\ 
 \\
\boldmath$< z > = 2.2$\unboldmath \\

MS &       29 &   2.24 &   0.99 & 181.86 &   1.46 &    10.8$\pm$1.3 &     0.52$\pm$0.03 &     0.61$\pm$0.05 \\ 

above MS &       26 &   2.28 &   1.25 & 571.21 &   4.24 &    29.3$\pm$3.3 &     0.67$\pm$0.02 &     0.51$\pm$0.05 \\ 

all &       55 &   2.27 &   1.15 & 367.32 &   2.75 &    19.5$\pm$2.2 &     0.59$\pm$0.02 &     0.56$\pm$0.04 \\ 
 \\
\boldmath$< z > = 4.4$\unboldmath \\

MS &        6 &   4.28 &   0.35 & 117.36 &   1.19 &     4.3$\pm$0.6 &     0.58$\pm$0.05 &     0.42$\pm$0.04 \\ 

above MS &        9 &   4.07 &   0.66 & 583.19 &   4.59 &    13.4$\pm$2.4 &     0.68$\pm$0.06 &     0.24$\pm$0.03 \\ 

all &       15 &   4.20 &   0.60 & 399.54 &   3.55 &    10.6$\pm$2.2 &     0.64$\pm$0.04 &     0.31$\pm$0.04 \\ 
 \\
\bf{ all z} \\

MS &       54 &   2.07 &   0.94 & 138.43 &   1.46 &     8.8$\pm$0.8 &     0.49$\pm$0.02 &     0.76$\pm$0.05 \\ 

above MS &       60 &   2.10 &   1.06 & 422.55 &   4.30 &    18.9$\pm$2.3 &     0.62$\pm$0.02 &     0.53$\pm$0.04 \\ 

all &      114 &   2.10 &   1.02 & 288.46 &   2.98 &    14.2$\pm$1.3 &     0.56$\pm$0.02 &     0.64$\pm$0.04 \\ 
 \\

  \\ 
\enddata\label{summary_table}
\tablenotetext{}{\noindent Samples include all galaxies with $M_{mol} > 10^{10}$\msun. Uncertainties in estimates of the means and the ratios were derived from bootstrap resampling of the data samples.}
\end{deluxetable*}

\section{Summary and Comments}\label{discuss}

We have provided a thorough physical and empirical foundation for the use of submm flux measurements as a probe
of the interstellar medium gas mass in galaxies (Appendix \ref{dust_app}). We find that a single empirical scaling 
exists between the specific luminosity of the RJ dust continuum ($L_{\nu_{850\mu m}}$) and the mass as determined from CO(1-0)
measurements over 3 orders of magnitude in $L_{\nu_{850\mu m}}$ (see Figure \ref{empir_cal}).\footnote{The SMG galaxies are probably 
gravitationally lensed so we do not include their luminosities in this estimate of the dynamic range.}

The ALMA Band 6 \& 7 observations with typically only a few minutes of integration detect a majority (79\%) of the 145 galaxies in our sample 
at z = 1 - 6 (see Figures \ref{detection_rates} and \ref{detection_rates1}). Thus, with ALMA, this technique immediately enables surveys of large numbers 
of objects. Using the RJ dust continuum, one also avoids the uncertainties of CO excitation variations which enter when translating 
higher rotational line measurements into equivalent CO(1-0) line luminosities and hence molecular gas masses. 

The appropriate temperature characterizing 
the RJ dust emission is a mass-weighted $T_d$ and, from basic understanding of the dust emission, it is clear that a luminosity-weighted $T_D$ determined from SED fitting
in each individual source should not be used in this technique -- rather it is more appropriate to simply adopt a constant value with $T_D \simeq 25$ K as done here.  
This statement applies to global measurements of the dust continuum, not instances where small regions of a galaxy are resolved and have 
locally enhanced mass-weighted $T_D$ \citep[e.g. the compact nuclei in Arp220,][]{sco15}.
The galaxies used in our empirical 
calibration and those observed by us using ALMA are all fairly massive ($M_{stellar} > 2\times10^{10}$ \msun) so we are not exploring low metallicity systems 
where the calibration may depart from constancy due to variations in the dust-to-gas abundance ratio or the dust properties. 
 
The results presented here suggest that:
\begin{itemize} 
\item At high redshift, the primary difference between galaxies with sSFR above the MS and those on the MS is simply  
increased gas contents of the former, not higher efficiency for conversion of gas to stars. 

However, 
\item the shorter ($\sim5\times$) gas depletion times at high redshift of {\bf all star forming galaxies, both on and above the MS}, imply a more efficient mode for star formation 
from existing gas supplies. This is naturally a result of highly dispersive gas motions (due to prodigious on-going 
accretion needed to replenish gas contents and to galaxy interactions) for all high redshift galaxies -- those on and above the MS. 
\end{itemize}

Our result of a single SF law at high redshift is very different from some prior studies. \cite{dad10} and \cite{gen10} obtain different SF laws for normal SF 
galaxies and starburst/SMG galaxies; however, in both cases, their ISM masses at high redshift are derived from higher-J CO transitions and they use different high-J to J = 0 line ratios and CO-to-H$_2$ conversion factors for the two classes of galaxies \citep[see also][]{sar14}. \cite{gen15}  compared CO and dust continuum results in order to constrain the variations in the conversion factor between the MS and starburst population, obtaining general agreement with their earlier results.  That their SF laws differ for normal and starburst/SMG galaxies is a result of their use of different CO-to-H$_2$ conversion factors, which we argue is inappropriate for global ISM measures (Appendix \ref{dust_app}). The technique developed by us here avoids the additional uncertainty introduced when observing higher J CO transitions at high redshift and global variations in the mass-weighted $T_D$ are likely to be small. 

\acknowledgments

We thank Zara Scoville for proof reading the manuscript and Sue Madden for suggesting use of the SPIRE 
fluxes for calibration. 
This paper makes use of the following ALMA data: ADS/JAO.ALMA\# 2013.1.00034.S. 
This paper makes use of the following ALMA data:
  ADS/JAO.ALMA\#2013.1.00111.S.  ALMA is a partnership of ESO (representing its member states), 
  NSF (USA) and NINS (Japan), together with NRC (Canada), NSC and ASIAA (Taiwan), and KASI 
  (Republic of Korea), in cooperation with the Republic of Chile. 
  The Joint ALMA Observatory is operated by ESO, AUI/NRAO and NAOJ.
			The National Radio Astronomy Observatory is a facility of the National
			Science Foundation operated under cooperative agreement by Associated
			Universities, Inc. RJI acknowledges support from ERC in the form of the Advanced Investigator Programme, 321302, COSMICISM.
\bibliography{scoville_dust}{}

\appendix

\section{Long Wavelength Dust Continuum as an ISM Mass Tracer}\label{dust_app}

Here, we summarize the physical and empirical basis for using long wavelength dust emission as a probe of ISM 
mass. The empirical calibration is obtained from: 1) a sample of 30 local star forming galaxies; 2) 12 low-z Ultraluminous Infrared Galaxies (ULIRGs);  and 3) 30 z $\sim$ 2 submm galaxies (SMGs). We have completely redone the analysis of these three galaxy
samples. We use Herschel SPIRE 350 and 500$\mu$m data which provide more reliable total submm fluxes for the extended objects at low redshift than 
were available from the SCUBA 850$\mu$m observations used in \cite{sco14}. 

The major differences, compared to the empirical calibration presented in \cite{sco14},
are that here we empirically calibrate the submm fluxes relative to the molecular gas masses rather than HI plus H$_2$ (since HI is not well measured in 
many ULIRGs and not at all in the z = 2 SMGs). We also go back to the original sources for the CO (1-0) luminosities and use a single CO to H$_2$ conversion 
factor for all objects; we expand the samples when we find additional global CO and SPIRE flux measurements and we remove a couple objects where we find errors from SINGs/KINGFISH surveys \citep{dra07b}. All of these calibrations yield a very similar rest frame 850$\mu$m luminosity per unit molecular gas mass with small dispersions. 
The mean value is also within 10\% of the value obtained by Planck for Milky Way molecular gas.

\subsection{Rayleigh-Jeans Dust Continuum -- Analytics}

The far infrared-submm emission from galaxies is dominated by dust re-emission of the luminosity from stars and active galactic nuclei (AGN). The luminosity at the peak of the FIR is often used to estimate the luminosity of obscured star formation or AGN. Equally important (but not often stressed) is the fact that the long-wavelength RJ tail of dust emission is nearly always optically 
thin, thus providing a direct probe of the total dust and, hence, the ISM mass -- provided the dust emissivity per unit mass  
and the dust-to-gas abundance ratio can be constrained. Here, we take the approach of {\bf empirically calibrating the appropriate combination 
of these quantities rather than requiring determination of each one independently.} 

The observed flux density from a source at luminosity distance d$_L$ is 
\begin{eqnarray}
S_{\nu_{obs}} &=&  { (1 - e^{- \tau_{d}(\nu_{rest})}) ~B_{\nu_{rest} }(T_d) (1+z) \over{ d_L^2}} 
 \end{eqnarray}

\noindent where B$_{\nu_{rest}}$ is the Planck 
function in the rest frame and $\tau_{d}(\nu)$ is the source optical depth at the emitted frequency. (The factor 1 + z accounts for 
the compression of frequency space in the observer's frame if the source is at significant redshift.) The source optical depth
is given by $\tau_{d}(\nu) = \kappa(\nu) \times 1.36 N_H m_H = \kappa(\nu) \times M_{gas}$, where $\kappa(\nu)$ is the absorption coefficient of the dust 
per unit {\it total mass of gas} (i.e. the effective area per unit mass of gas), N$_H$ is the column density of H nuclei and the factor 1.36 accounts for the mass contribution of heavier atoms 
(mostly He at 8\% by number). Often the dust opacity coefficient is specified per unit mass of dust. However, here we are empirically calibrating the dust opacity relative to the ISM molecular gas mass, so it is convenient to use the above definition, avoiding a separate specification of the dust opacity coefficient per mass of dust and the dust-to-gas ratio.

At long wavelengths where the dust is optically thin, the flux density is then  
\begin{equation}
S_{\nu_{obs}} =  {M_{mol} \kappa(\nu_{rest}) B_{\nu_{rest} } (1+z) \over{ d_L^2}} .
\label{fnu_1}
\end{equation}

\noindent Because $\kappa$ is per unit gas mass, the gas-to-dust ratio is absorbed in $\kappa$ and it therefore does not appear explicitly in Equation A2.
Written with a Rayleigh-Jeans $\nu ^2$ dependence, appropriate at long wavelengths, Equation A2 becomes 
\begin{equation}
S_{\nu_{obs}} =  {M_{mol} \kappa(\nu_{rest})  2kT_{\rm d} (\nu_{rest}/c)^2 {\it{\Gamma}_{RJ}(T_d,\nu_{obs},z) (1+z)}\over{d_L^2}} 
\label{fnu}
\end{equation}

\noindent where $\Gamma_{\rm RJ}$ is the correction for departure in the rest frame of the Planck function from Rayleigh-Jeans (i.e. $B_{\nu_{rest}} / RJ_{\nu_{rest}}$).  $\Gamma_{\rm RJ}$  is given by 
\begin{eqnarray}
\it{\Gamma}_{\rm RJ}(T_d,\nu_{obs}, z) &=&  {h \nu_{obs} (1+z) / k T_d \over{e^{h \nu_{obs} (1+z) / k T_d} -1 }} ~.
  \end{eqnarray}
  
  Equation \ref{fnu} can be rewritten for the specific luminosity ($L_{\nu_{rest}}$) in the rest frame of the galaxy, 
\begin{eqnarray}
L_{\nu_{rest}}  &=& S_{\nu_{obs}} 4\pi d_L^2 / (1+z)  \nonumber \\
              &=&   \kappa(\nu_{rest})  8\pi kT_{d} (\nu_{rest}/c)^2 \it{\Gamma}_{RJ} M_{mol} ~.
\end{eqnarray}

The long wavelength dust opacity can be approximated by a power-law in wavelength:
\begin{equation}
\kappa (\nu)  = \kappa (\nu_{850\mu \rm m}) (\lambda/850\mu \rm m)^{-\beta} .
\end{equation}\label{kappa}We adopt $\lambda = 850\mu$m ($\nu = 353$ GHz) as the fiducial wavelength
since it corresponds to most of the high z SCUBA observations and is the optimum for ALMA (i.e. Band 7). We will use a spectral index $\beta \simeq 1.8$ (see Section  \ref{beta}). 

The rest frame luminosity-to-mass ratio at the fiducial wavelength is given by 
\begin{eqnarray}
{L_{\nu_{850\mu m}} \over {M_{mol}}} &=&  \kappa (\nu_{850\mu m})  {{8\pi k {\nu^2}}\over{c^2}} T_{ d} \it{\Gamma_{RJ}}  ~~~~~\rm and ~ we ~ define~ \nonumber \\
 \alpha_{\nu_{850\mu m}} &\equiv& { L_{\nu_{850\mu m}} \over{ M_{mol}}}  = {{8\pi k {\nu^2}}\over{c^2}} \kappa (\nu_{850\mu m}) T_{d}  \it{\Gamma_{RJ}} ~.
\end{eqnarray}
\noindent In Section \ref{empirical} we show that this luminosity-to-mass ratio ($ \alpha_{\nu_{850\mu m}}$) is relatively constant under a wide range of conditions 
in normal star-forming and starburst galaxies and at both low and high redshift. Then, once this constant is empirically calibrated, we use measurements 
of the RJ flux density and, hence the luminosity to estimate gas masses. \bf We note that this result is equivalent to a constant molecular gas mass to dust mass ratio over a wide range of redshifts
for high stellar mass galaxies. \rm

\subsection{Mass-weighted T$_d$}\label{temp}

It is important to recognize that the dust temperature relevant to the RJ emission tail is a {\bf mass-weighted} $<T_d>_{M}$. This is definitely not the same 
as the luminosity-weighted $<T_d>_{L}$ which might be derived by fitting the IR SED (determined largely by the wavelength of peak IR emission). 
The former is a linear weighting with $T_d$; the latter is weighted as $\sim T^{5-6}$ depending on $\beta$. In dust clouds with temperature 
gradients, these temperatures are likely to differ by a factor of a few (depending on the optical depths and mass distributions).  

In local star-forming galaxies, the mass-weighted $<T_d>_{M} \sim 15 -- 35$K,  
and even in the most vigorous starbursts like Arp 220, the mass-weighted dust temperature is probably less than 45 K if one considers the entire galaxy; in contrast, the luminosity-weighted 
$<T_d>_{M} \sim 50 - 200$ K for Arp 220, depending on the size of the region. It is incorrect then to do an SED fit and use the derived temperature for estimation of the masses. 

In fact, variations 
in the effective dust temperature are probably small on galactic scales since theoretically one expects that the mass-weighted $<T_d>_M$ should depend on the $\sim 1/6$'th 
power of mean radiation energy density. The observed submm fluxes therefore directly probe the total mass of dust and depend only linearly on $<T_d>_{M}$ which varies very little.
\cite{mag14} investigated the variations in $T_D$ derived from SED fits for stacks of galaxies at z = 1 to 2.3 from the MS to a factor 10 above the MS. The 
temperatures were found to increase from $\sim25$ to 33 K going to sSFR 10 times above the MS. Once again, we emphasize that those $T_D$ are luminosity-weighted, not mass-weighted,
but even so, they do not indicate very large variations. \cite{gen15} have also advocated the use of a variable $T_D$ to reduce apparent scatter in the 
relationship between high J CO lines and the dust continuum; however, much of this scatter is likely due to CO excitation variations which enter from use of higher J transitions, so we do not follow this route. 

 In practice, it will be difficult or impossible 
to determine $<T_d>_{M}$ in most sources since the observed SEDs are not of sufficient accuracy to measure the small secondary peak due to the cold dust on the 
RJ tail of the SED. Moreover, this peak is unlikely to be discrete since there will be a range of temperatures in the cold component. In the Galaxy, the Planck data show 
$<T_d>_{M} = 15 - 22$ K \citep{pla11a}. Recognizing that most of the galaxies with higher SF rates at high redshift are likely to have slightly elevated dust temperatures, 
we adopt \boldmath$\rm <T_d>_M = 25~\rm K$\unboldmath ~for numerical estimates when necessary and might reasonably expect a range of 20 - 35 K. Since the mass estimates vary as $T_d^{-1}$, this 
range implies less than 25 - 30\% variation associated with the expected range of {\bf global} mass-weighted dust temperatures.  

\begin{figure}[ht]
\epsscale{0.5}  
\plotone{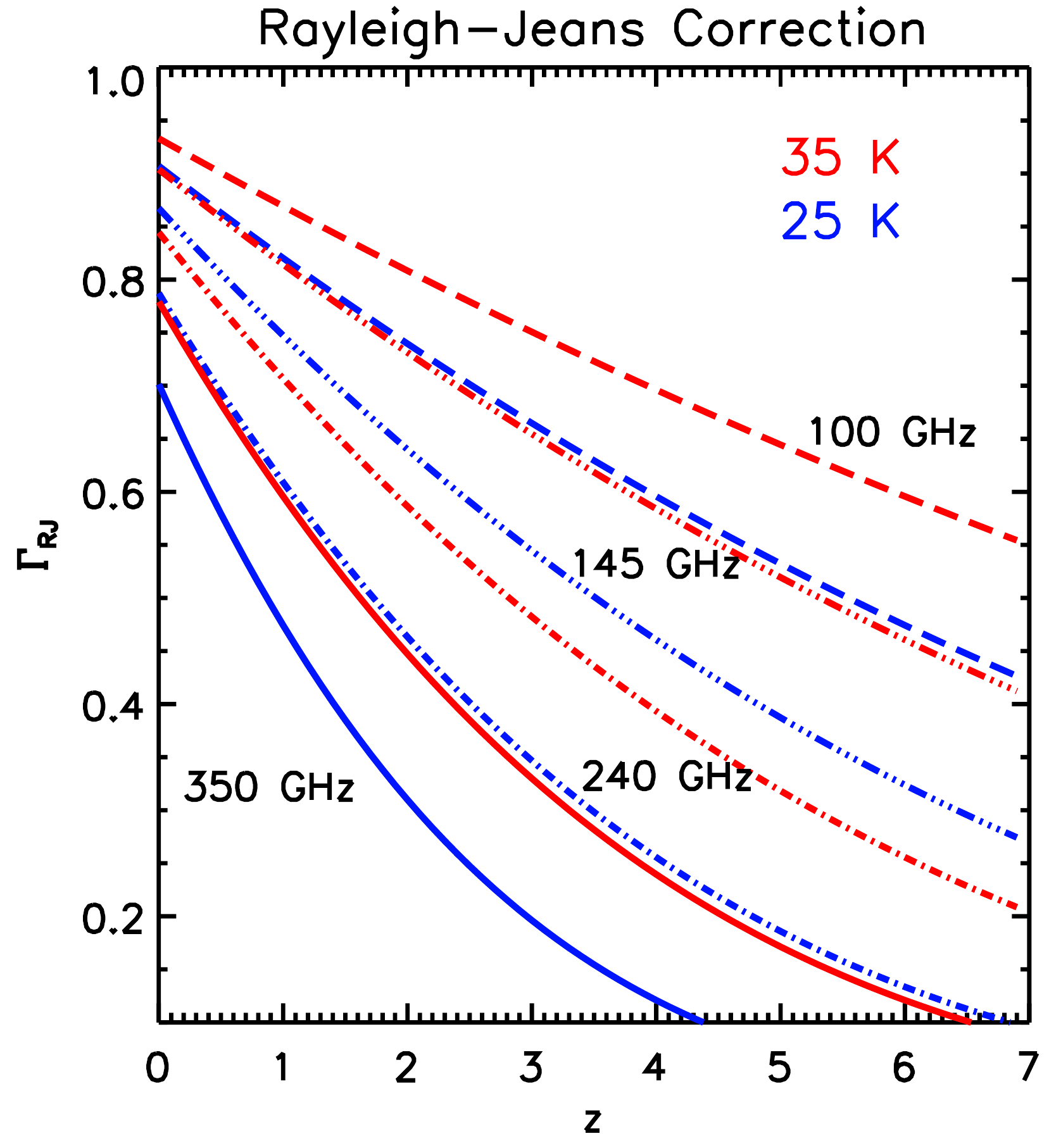}
\caption{The RJ correction factor $\Gamma_{RJ}$ is shown as a function of redshift for 4 ALMA Bands and dust temperatures of 25 and 35 K. }
\label{rj} 
\end{figure}

Figure \ref{rj} shows the  $\Gamma_{RJ}$ correction factor for dust temperatures of 25 and 35 K. [$\Gamma_0$ is the value of $\Gamma$ appropriate to 
the z = 0, $T_d=25$ K and $\lambda= 850\mu \rm m$ used to calibrate $\alpha_{850\mu \rm m}$; $\Gamma_0 = 0.71$ (see Fig. \ref{rj}).] 

\subsection{The Dust Submm Spectral Index -- $\beta$}\label{beta}

In order to relate submm flux measurements of galaxies at different redshifts (i.e. different rest frame wavelengths), to an empirically constrained mass-light ratio 
at rest frame 850$\mu$m, one needs to know the spectral index of the RJ dust emission. 
The overall spectral slope of the rest frame submm dust emission flux density (Equation \ref{fnu}) is observed to
vary as $S_{\nu} \propto \nu^{\alpha}$ with $\alpha = 3 - 4$. Two powers of $\nu$ are from the RJ dependence;  the 
remainder is due to the frequency variation in $\kappa(\nu) \propto \nu^{\beta}$. 
Most theoretical models for the dust have opacity spectral indices of $\beta$ = 1.5 -- 2 \citep{dra11}. Empirical fits to the observed long wavelength
SEDs suggest $\beta$ = 1.5 -- 2 (Dunne \& Eales 2001, Clements, 
Dunne \& Eales 2009) for local galaxies. Probably the best determination at high redshift is that of \cite{cha09} who used their $\lambda =1.1$ survey to find
$< \beta > = 1.75$ for 29 SMGs with a median z = 2.7.

Planck has provided a robust Galactic determination of $\beta$ using 7 submm bands (at $\lambda$ = 3 mm to 100 $\mu$m). For the Taurus 
cloud complex, $\beta = 1.78 \pm 0.08$ for both atomic and molecular ISM regions \citep{pla11a}.  And for the Galaxy, \cite{pla11b} finds $\beta = 1.8\pm 0.1$ with no significant difference between the HI and H$_2$-dominant regions. We therefore adopt \boldmath${\beta = 1.8}$\unboldmath ~ when needed in the analysis below. 

\subsection{Empirical Calibration from Local Galaxies, Low-z ULIRGs and z = 2 SMGs}\label{empirical}

In order to empirically calibrate the 850$\mu$m dust opacity per unit gas mass we make use of galaxy samples for which 
both the submm dust emission and molecular gas masses are well-determined globally for the whole galaxy. 
Our local sample includes 28 star forming galaxies from the Herschel KINGFISH survey \citep{dal12} and 12 ULIRGs from the Herschel VNGS and GOALS surveys \citep[][Chu \etal in prep.]{coo12}. These have full imaging in the SPIRE 350 and 500$\mu$m bands.\footnote{These SPIRE data are far superior to the earlier SCUBA imaging for extended galaxies. The ground-based SCUBA observations, taken in beam chopping mode to remove atmospheric background, can cancel extended emission components.}

For estimation of the associated gas masses we have used exclusively CO (1-0) data for which there are also global CO luminosity measures \citep{you95,san91,sol97,san89} with consistent single-dish calibrations. Although some of these galaxies have mm-interferometric imaging, those data often resolve out larger spatial components and therefore 
often recover less than 50\% of the single dish line fluxes. For the high redshift SMGs, there are no HI measurements so we restrict our empirical calibration 
entirely to molecular gas masses. 

At high redshift, we make use of a sample of 30 SMGs for which there exist good SNR measurements of CO (1-0) from JVLA. For this 
sample we use SCUBA 850$\mu$m fluxes since the longer wavelength (compared to SPIRE 500$\mu$m) is needed to stay on the RJ tail and the sources are also 
quite compact. Most of these objects are at z $< 2.5$. Based on the high submm fluxes, it is clear that many of these SMGs are strongly lensed. This means 
that the \it{apparent} \rm $L_{\nu_{850\mu \rm m}}$ and M$_{mol}$ are large over-estimates of their true values. However, it is reasonable to assume that the magnifications are similar 
for the dust and the gas emission since they both arise in cold ISM. Thus, the $L_{\nu_{850\mu \rm m}}/M_{mol}$ will provide a consistency check (at high SNR) as to whether 
the relevant combination of the dust-to-gas mass ratio, the dust opacity function and the mass-weighted $<T_d>_M$ are similar to that in the low z calibrators. 

Our  restriction to calibration samples with good CO (1-0) line measurements is very important. Only the CO(1-0) line luminosities have 
been well correlated with virial masses from large Galactic samples of self-gravitating GMCs \citep{sco87,sol87}. The higher CO transitions have excitation-dependent 
flux ratios relative to the 1-0 emission luminosities both in Galactic GMCs \citep{san93} and in high z galaxies \citep[][see Figure 4]{car13}. For high redshift galaxies, this necessarily restricts calibration samples to those observed at high signal-to-noise ratio with JVLA or possibly GBT. 

The observed SPIRE 500$\mu$m (local galaxies and ULIRGs: Tables \ref{tab:local_gal} and \ref{tab:ulirg}) and SCUBA 850$\mu$m (SMGs: Table \ref{tab:smg}) fluxes were converted to $850\mu$m specific luminosity $L_{\nu(850\mu \rm m)}$ of the assumed 25K dust using
\begin{eqnarray}
L_{\nu(850\mu \rm m)} & =& 1.19\times10^{27}~S_{\nu}[{\rm Jy}]~{\left(\nu(850\mu \rm m) \over{\nu_{obs} (1+z)}\right)^{3.8}} ~~{(\it d_{L}[\rm Mpc])^2 \over {1+z}} ~~
{\it \Gamma_{RJ}(25,\nu_{850\mu m},0) \over{\it \Gamma_{RJ}(25,\nu_{obs},z)}}~\rm [ergs~ sec^{-1} Hz^{-1}] .
\label{lnu_eqn}
\end{eqnarray}
\noindent (The term with ratios of $\Gamma_{RJ}$ is necessary since we wish to estimate $L_{\nu(850\mu \rm m)}$ which would be associated with 25K dust producing the same $\rm L_{\nu}$ at $\nu=\nu_{obs}\times(1+z)$ in the observed galaxy rest frame.)

Figure \ref{empir_cal}-Left shows the CO(1-0) luminosities for the three samples of galaxies plotted as a function of their  specific continuum luminosities at $\lambda = 850\mu$m. All three sets of these very diverse galaxies (normal star-forming, ultraluminous starbursts and high redshift starbursts) fall on the same  1:1 line 
and this provides the empirical basis for using the RJ continuum as a tracer of ISM molecular gas mass. We note that many of these SMGs are likely lensed, but the high magnifications allow extension of the calibration to high redshift with excellent signal-to-noise ratios. The dust continuum 
and CO(1-0) emission experiences the same lensing magnifications since the SMGs lie on the same linear correlation as the unmagnified objects at low redshift.

Molecular gas masses were computed from the CO(1-0) integrated fluxes (S$\Delta v$) or line luminosities 
(L$^{\prime}_{CO}$) using the relations \citep{sol05,bol13}:

\begin{eqnarray}
L^{\prime}_{CO}[\rm K~km~s^{-1} pc^2]&=&3.25\times10^{7}~(S\Delta v[\rm Jy~km~s^{-1}]) (\nu_{rest}[GHz])^{-2} (1+z)^{-1} (\it{d}_{L}[Mpc])^2  \\
M_{mol}[\msun] &=& 6.5~ L^{\prime}_{CO}[\rm K~km~s^{-1} pc^2]  .  \label{co}
\end{eqnarray}

The constant ($\alpha_{CO} = 6.5 ~\msun / \rm{K~km~s^{-1} pc^2}$) in Equation A10 is based on a standard Galactic conversion factor $X_{CO} = 3\times10^{20}$ N(H$_2$) cm$^{-2}$ (K km s$^{-1})^{-1}$ (see below) and 
it includes a factor 1.36 to account for the associated mass of heavy elements (mostly He at 8\% by number). For the local SF galaxies, the integrated 
CO (1-0) fluxes were all taken from \cite{you95} so they have consistent calibration and technique for integrating over extended galaxies. (In the course of this work, we found that the molecular 
gas masses used in the KINGFISH papers \citep{dra07b,dal12} are actually based on the same \cite{you95} CO survey even though they reference 
\cite{ken03} for the molecular masses. \cite{dra07b} argues for (and used) a higher $X_{CO}= 4\times10^{20}$, so their H$_2$ masses are larger (they don't explicitly include the He contribution).

We note that \cite{bol13} have advocated a value of $X_{CO}=2\times10^{20}$ cm$^{-2}$ (K km s$^{-1})^{-1}$ based largely 
on Galactic $\gamma$-ray emission analysis. However, this relies on low angular resolution $\gamma$-ray (CosB, Egret, CGRAO  and Fermi) and CO (Columbia Survey) datasets which highly 
weights gas in the solar neighborhood (mostly HI), rather than the molecular ring in the inner galaxy; it also does not resolve the distant GMCs. The $\gamma$-ray approach also relies on the 
questionable assumption that the high energy cosmic rays which produce the $\gamma$-rays are approximately constant 
in the Galactic disk, and that these particles penetrate fully the dense molecular clouds. It is noteworthy that there are also large variations in the $\gamma$-ray-based X$_{CO}$ values 
between the different analyses and as a function of Galactic radius \citep{bol13}.  In contrast, extensive GMC surveys (with samples of more than 500 resolved GMCs in the inner Galactic plane, independently 
measured and analyzed) yielded X$_{CO} = 
3.6$ and 3.0$\times10^{20}$ cm$^{-2}$ (K km s$^{-1})^{-1}$ \cite[respectively:][corrected to $R_0 = 8.5$ kpc]{sco87,sol87}. 

We have chosen to use a single conversion factor ($\alpha_{CO}$ or X$_{CO}$) for all galaxies. A several times smaller conversion factor is often used for ULIRGs and SMGs in analyzing the CO transitions seen from the hot, dense nuclear regions of these merger systems.  Similar to the argument we have given above for a uniform T$_D$, a smaller value of $\alpha_{CO}$ is inappropriate for the globally distributed molecular gas although it will sometimes be appropriate for high resolution observations which isolate the nuclear regions.  

\begin{deluxetable*}{lrrcrrrrcc}[ht]
\tablecaption{Low-$z$ Star Forming Galaxies with global CO(1-0) \& Herschel SPIRE Luminosities}
\tablewidth{0pt}
\tablehead{ \colhead{Galaxy} & \colhead{Distance} & \colhead{S$_{CO(1-0)}$} & \colhead{log L$^{\prime}_{CO(1-0)}$} & \colhead{log M$_{mol}$} & \colhead{S$_{\nu}350\mu m$} & \colhead{S$_{\nu}500\mu m$}  & \colhead{S$_{\nu}850\mu m$} & \colhead{L$_{\nu_{850\mu \rm m}} $} & \colhead{L$_{\nu_{850\mu \rm m}} /M_{\rm mol}$}  \\
\colhead{}  & \colhead{Mpc} & \colhead{Jy} & \colhead{K km s$^{-1}$ pc$^2$ }  &  \colhead{\msun} &  \colhead{Jy} &  \colhead{Jy}&  \colhead{Jy}  & \colhead{$10^{30}$ cgs} & \colhead{$10^{20}$ cgs/\msun}}
\startdata
    Antennae &         21.4 &         2000 &         9.35 &        10.16 &        14.06 &         4.69 &        0.803 &        0.440 &         0.30 \\
       IC342 &          3.4 &        29220 &         8.92 &         9.73 &       247.95 &        96.90 &       16.886 &        0.233 &         0.44 \\
     NGC0628 &         11.4 &         2160 &         8.83 &         9.65 &        29.07 &        12.64 &        2.185 &        0.340 &         0.76 \\
     NGC1482 &         22.0 &          560 &         8.82 &         9.63 &         6.03 &         2.10 &        0.359 &        0.208 &         0.48 \\
     NGC2146 &         15.0 &         2840 &         9.19 &        10.01 &        22.13 &         7.08 &        1.220 &        0.328 &         0.32 \\
     NGC2798 &         24.7 &          440 &         8.82 &         9.63 &         2.76 &         1.03 &        0.175 &        0.128 &         0.30 \\
     NGC2841 &          9.8 &         1870 &         8.64 &         9.45 &        15.20 &         6.66 &        1.153 &        0.132 &         0.47 \\
     NGC2976 &          3.6 &          610 &         7.27 &         8.09 &        11.11 &         4.55 &        0.793 &        0.012 &         0.98 \\
     NGC3184 &          8.1 &         1120 &         8.25 &         9.07 &        14.53 &         6.39 &        1.109 &        0.087 &         0.75 \\
     NGC3351 &          9.3 &          700 &         8.17 &         8.98 &        13.01 &         5.05 &        0.876 &        0.091 &         0.95 \\
     NGC3521 &          9.0 &         4920 &         8.99 &         9.80 &        44.84 &        18.43 &        3.194 &        0.309 &         0.49 \\
     NGC3627 &          8.9 &         4660 &         8.95 &         9.77 &        35.72 &        13.68 &        2.371 &        0.225 &         0.38 \\
     NGC3938 &         14.0 &         1750 &         8.92 &         9.73 &         9.78 &         4.12 &        0.711 &        0.167 &         0.31 \\
     NGC4254 &         20.0 &         3000 &         9.47 &        10.28 &        25.27 &         8.70 &        1.493 &        0.714 &         0.38 \\
     NGC4321 &         20.0 &         3340 &         9.51 &        10.33 &        26.50 &        10.26 &        1.760 &        0.842 &         0.40 \\
     NGC4536 &         25.0 &          740 &         9.05 &         9.86 &        11.97 &         5.25 &        0.897 &        0.670 &         0.92 \\
     NGC4569 &         20.0 &         1500 &         9.16 &         9.98 &         8.94 &         3.49 &        0.598 &        0.286 &         0.30 \\
     NGC4579 &         20.0 &          910 &         8.95 &         9.76 &         8.43 &         3.36 &        0.577 &        0.276 &         0.48 \\
     NGC4631 &          9.0 &         1740 &         8.54 &         9.35 &        51.77 &        22.80 &        3.952 &        0.383 &         1.72 \\
     NGC4725 &         17.1 &         1950 &         9.14 &         9.96 &        15.77 &         7.53 &        1.296 &        0.453 &         0.50 \\
     NGC4736 &          5.3 &         2560 &         8.24 &         9.06 &        26.60 &        11.21 &        1.950 &        0.066 &         0.58 \\
     NGC4826 &          5.6 &         2170 &         8.22 &         9.03 &        15.58 &         5.99 &        1.041 &        0.039 &         0.36 \\
     NGC5055 &          8.2 &         5670 &         8.97 &         9.78 &        60.99 &        24.80 &        4.301 &        0.346 &         0.57 \\
     NGC5194 &          8.2 &         9210 &         9.18 &         9.99 &        62.60 &        21.28 &        3.691 &        0.297 &         0.30 \\
     NGC5713 &         26.6 &          680 &         9.07 &         9.88 &         6.07 &         2.18 &        0.372 &        0.315 &         0.41 \\
     NGC5866 &         12.5 &          250 &         7.98 &         8.79 &         2.98 &         1.08 &        0.187 &        0.035 &         0.57 \\
     NGC6946 &          5.5 &        12370 &         8.96 &         9.77 &       103.55 &        40.66 &        7.071 &        0.256 &         0.43 \\
     NGC7331 &         14.7 &         4160 &         9.34 &        10.15 &        38.57 &        15.68 &        2.702 &        0.700 &         0.49 \\
 \enddata
\tablecomments{CO (1-0) line fluxes from \cite{you95}; SPIRE 350 and 500$\mu$m fluxes from \cite{dal12} (except NGC 5194 -- Herschel VNGS data release Wilson \etal) with a multiplicative factor of 0.95 for extended source color correction \citep[see][]{rem15}. The 850$\mu$m fluxes listed in column 8 were extrapolated from the 500$\mu$m fluxes assuming a spectral index $\beta=1.8$ and $\Gamma_{RJ}$ appropriate for $T_d$ = 25K.  L$_{\nu_{850\mu \rm m}}$ was computed from the observed fluxes using Equation \ref{lnu_eqn}.}\label{tab:local_gal}
\end{deluxetable*}

\begin{deluxetable*}{lrrcrrrrcc}[ht]
\tablecaption{ULIRG Galaxies with global CO(1-0) \& Herschel SPIRE Luminosities}
\tablewidth{0pt}
\tablehead{ \colhead{Galaxy} & \colhead{Distance}  & \colhead{log L$^{\prime}_{CO(1-0)}$} & \colhead{log M$_{mol}$} & \colhead{CO ref.} & \colhead{S$_{\nu}350\mu m$} & \colhead{S$_{\nu}500\mu m$}  & \colhead{S$_{\nu}850\mu m$} & \colhead{L$_{\nu_{850\mu \rm m}} $} & \colhead{L$_{\nu_{850\mu \rm m}} /M_{\rm mol}$}  \\
\colhead{}  & \colhead{Mpc}  & \colhead{K km s$^{-1}$ pc$^2$ }  &  \colhead{\msun} &  \colhead{} &  \colhead{Jy} &  \colhead{Jy}&  \colhead{Jy}  & \colhead{$10^{30}$ cgs} & \colhead{$10^{20}$ cgs/\msun}}
\startdata
      1ZW107 &        170.0 &         9.62 &        10.44 &            1 &         0.72 &         0.20 &        0.029 &        1.014 &         0.37 \\
      Arp148 &        143.0 &         9.54 &        10.35 &            1 &         0.92 &         0.27 &        0.040 &        0.988 &         0.44 \\
      Arp220 &         79.0 &         9.76 &        10.57 &          1,2 &        10.89 &         3.60 &        0.584 &        4.359 &         1.16 \\
IRASF05189-2 &        168.0 &         9.72 &        10.53 &            1 &         0.59 &         0.16 &        0.023 &        0.790 &         0.23 \\
IRASF08572+3 &        232.0 &         9.13 &         9.94 &          3,2 &         0.16 &         0.05 &        0.007 &        0.429 &         0.49 \\
IRASF10565+2 &        176.0 &         9.73 &        10.54 &          1,2 &         1.15 &         0.32 &        0.047 &        1.749 &         0.50 \\
IRASF12112+0 &        292.0 &         9.96 &        10.77 &            1 &         0.65 &         0.19 &        0.024 &        2.489 &         0.42 \\
IRASF14348-1 &        330.0 &        10.12 &        10.93 &            1 &         0.61 &         0.17 &        0.022 &        2.834 &         0.33 \\
IRASF22491-1 &        301.0 &         9.77 &        10.58 &            1 &         0.24 &         0.06 &        0.008 &        0.866 &         0.23 \\
      Mrk231 &        174.0 &         9.72 &        10.53 &          1,2 &         1.83 &         0.51 &        0.076 &        2.741 &         0.80 \\
      Mrk273 &        153.0 &         9.67 &        10.49 &          2,3 &         1.36 &         0.37 &        0.055 &        1.551 &         0.51 \\
\enddata
\tablecomments{CO (1-0) line fluxes from:  \cite[1][]{san91}, \cite[2][]{sol97} and \cite[3][]{san89}; SPIRE 350 and 500$\mu$m fluxes from Chu \etal (2015, in prep.) for all galaxies except Arp 220 (Herschel VNGS data release Wilson \etal). L$_{\nu_{850\mu \rm m}}$ was computed from the observed fluxes using Equation \ref{lnu_eqn}.}\label{tab:ulirg}
\end{deluxetable*}

\begin{deluxetable*}{lrccccrrcrcc}[ht]
\tablecaption{SMGs at z $\sim$ 2 with CO(1-0) data}
\tablewidth{0pt}
\tablehead{ \colhead{Galaxy} & \colhead{z}   & \colhead{Distance}  & \colhead{ref.} & \colhead{log L$^{\prime}_{CO(1-0)}$} & \colhead{log M$_{mol}$}  & \colhead{S$_{\nu}(\lambda\mu m)$} & \colhead{L$_{\nu_{850\mu \rm m}} $} & \colhead{L$_{\nu_{850\mu \rm m}} /M_{\rm mol}$}  \\
\colhead{}  & \colhead{}  & \colhead{Gpc}  & \colhead{} & \colhead{K km s$^{-1}$ pc$^2$ }  &  \colhead{\msun}  &  \colhead{mJy}   & \colhead{$10^{30}$ cgs} & \colhead{$10^{20}$ cgs/\msun}}
\startdata
%\\
EROJ164502+4 &         1.44 &         10.6 &            8 &        10.83 &        11.65 &    4.89(850) &       16.362 &          0.37 \\
H-ATLASJ0903 &         2.31 &         18.9 &            1 &        11.42 &        12.24 &   54.70(880) &      214.455 &          1.24 \\
H-ATLASJ0913 &         2.63 &         22.2 &            1 &        11.40 &        12.21 &   36.70(880) &      145.710 &          0.89 \\
H-ATLASJ0918 &         2.58 &         21.7 &            1 &        11.53 &        12.34 &   18.80(880) &       74.504 &          0.34 \\
H-ATLASJ1132 &         2.58 &         21.7 &            1 &        11.33 &        12.14 &  106.00(500) &      179.479 &          1.30 \\
H-ATLASJ1158 &         2.19 &         17.8 &            1 &        11.26 &        12.07 &  107.00(500) &      150.507 &          1.29 \\
H-ATLASJ1336 &         2.20 &         17.9 &            1 &        11.36 &        12.17 &   36.80(880) &      143.699 &          0.97 \\
H-ATLASJ1344 &         2.30 &         18.9 &            1 &        11.86 &        12.67 &   73.10(880) &      286.544 &          0.61 \\
H-ATLASJ1413 &         2.48 &         20.7 &            1 &        11.65 &        12.46 &   33.30(880) &      131.429 &          0.46 \\
HATLASJ08493 &         2.41 &         20.0 &           11 &        10.36 &        11.17 &    4.60(870) &       17.642 &          1.19 \\
HATLASJ08493 &         2.42 &         20.1 &           11 &        10.22 &        11.03 &    6.90(870) &       26.468 &          2.48 \\
HATLASJ08493 &         2.41 &         20.0 &           11 &        11.21 &        12.02 &   19.00(870) &       72.855 &          0.70 \\
HATLASJ08493 &         2.41 &         20.0 &           11 &        11.15 &        11.96 &   25.00(870) &       95.851 &          1.05 \\
     HLSW-01 &         2.96 &         25.6 &            2 &        11.66 &        12.48 &   52.80(880) &      212.973 &          0.71 \\
      HXMM01 &         2.31 &         19.0 &           12 &        11.66 &        12.48 &   27.00(880) &      105.868 &          0.35 \\
SMMJ02399-01 &         2.81 &         24.1 &            9 &        11.35 &        12.16 &   23.00(850) &       85.752 &          0.59 \\
SMMJ04135+10 &         2.85 &         24.5 &            6 &        11.39 &        12.20 &   25.00(850) &       93.462 &          0.59 \\
SMMJ04431+02 &         2.51 &         21.0 &           10 &        10.90 &        11.72 &    7.20(850) &       26.346 &          0.51 \\
SMMJ123549.4 &         2.20 &         17.9 &            6 &        10.89 &        11.71 &    8.30(850) &       29.839 &          0.59 \\
SMMJ123707.2 &         2.49 &         20.8 &            6 &        11.44 &        12.25 &   10.70(850) &       39.105 &          0.22 \\
SMMJ14009+02 &         2.93 &         25.3 &            9 &        11.09 &        11.91 &   15.60(850) &       58.678 &          0.73 \\
SMMJ14011+02 &         2.57 &         21.6 &           10 &        11.11 &        11.92 &   12.30(850) &       45.161 &          0.54 \\
SMMJ163550.9 &         2.52 &         21.1 &            9 &        10.97 &        11.78 &    8.40(850) &       30.754 &          0.51 \\
SMMJ163554.2 &         2.52 &         21.1 &            9 &        11.09 &        11.90 &   15.90(850) &       58.212 &          0.72 \\
SMMJ163555.2 &         2.52 &         21.1 &            9 &        10.83 &        11.65 &   12.50(850) &       45.765 &          1.04 \\
SMMJ163650.4 &         2.38 &         19.7 &            6 &        10.98 &        11.79 &    8.20(850) &       29.793 &          0.48 \\
SMMJ163658.1 &         2.45 &         20.4 &            6 &        11.04 &        11.85 &   10.70(850) &       39.022 &          0.55 \\
SMMJ2135-010 &         2.33 &         19.2 &            4 &        11.78 &        12.60 &  106.00(870) &      404.977 &          1.03 \\
SPT-S053816- &         2.79 &         23.8 &            5 &        11.64 &        12.46 &  125.00(870) &      488.049 &          1.71 \\
SPT-S233227- &         2.73 &         23.2 &            5 &        11.78 &        12.59 &  150.00(870) &      583.833 &          1.49 \\
\enddata
\tablecomments{Submm and CO (1-0) line fluxes from:  \cite[1][]{har12}, \cite[2][]{rie11}, \cite[3][]{les11}, 
\cite[4][]{tho15}, \cite[5][]{ara13}, \cite[6][]{ivi11}, \cite[7][]{car11} , \cite[8][]{gre03},\cite[9][]{tho12},
\cite[10][]{har10},\cite[11][]{ivi13} and \cite[12][]{fu13},. L$_{\nu_{850\mu \rm m}}$ was computed from the observed fluxes using Equation \ref{lnu_eqn}.}\label{tab:smg}
\end{deluxetable*}

\vfill
\eject

In Figure \ref{empir_cal} the ratios L$_{\nu_{850\mu \rm m}} /\rm {M_{ mol}}$ are plotted as a function of L$_{\nu_{850\mu \rm m}}$ for the three samples 
of galaxies listed in Tables \ref{tab:local_gal} - \ref{tab:smg}. The galaxies in all three samples clearly overlap in the luminosity-to-mass ratios 
and their mean ratios are indeed very similar. The mean of the local star-forming galaxies, ULIRGs and SMGs is 

\boldmath
\begin{eqnarray}
\alpha_{\nu} \equiv < L_{\nu_{850\mu \rm m}} /M_{\rm mol}> = 6.7\pm1.7\times 10^{19} \rm erg ~sec^{-1} Hz^{-1} {\msun}^{-1} 
\end{eqnarray}\label{alpha}
\unboldmath

and we adopt this value in the analysis below.

\subsection{Planck Measurements for HI and H$_2$ in the Galaxy}\label{pla}

The Planck measurements of the submm emission from the Galaxy provide both very high photometric 
accuracy and the ability to probe variations in the opacity to mass ratio between atomic and molecular phases, 
and with Galactic radius. (The latter could possibly provide a probe of metallicity dependence.)

In the Taurus complex, the \cite{pla11b} obtained resolved observations of the HI and H$_2$ ISM components with 
best fit  ratios of $\tau_{250\mu \rm m} / N_{\rm H} = 1.1 \pm 0.2 ~\rm ~and~ 2.32 \pm 0.3 \times 10^{-25}$ cm$^{2}$ for the 
atomic and molecular phases. 
The HI column densities were derived from the optically thin 21cm emission with a small  correction of 25\% for
optically thick 21 cm emission. The H$_2$ column densities were taken from \cite{pin10} who used NIR extinction measures as a primary measure 
of molecular gas column densities. (CO column 
densities were also obtained from a non-LTE radiative transfer analysis but these were not used for the Planck analysis). The mean dust temperature from the Planck observations was 18K derived  
in Taurus and the mean $<\beta> = 1.8$. We translate the value given above for $\tau_{250\mu \rm m} / N_{\rm H} $ in the molecular phase into 
a specific luminosity per unit mass of ISM (using $M_{\rm mol}$ = 1.36 $M_{\rm H_2}$ to account to He):

\begin{eqnarray}
 {L_{\nu_{850\mu \rm m}} \over M_{\rm H_2}}  &=&  \left[ \tau_{250\mu \rm m} / N_{\rm H} \right] \left({\nu_{850\mu \rm m} \over \nu_{250\mu \rm m}}\right)^{\beta} {4 \pi B_{\nu} (T_d) \over m_H} \nonumber \\
 &=& 8.4\times 10^{19}  \rm ergs/sec/Hz/\msun   \nonumber \\
  \alpha_{850\mu \rm m} &=& {L_{\nu_{850\mu \rm m}} \over M_{\rm mol}}  = 6.2\times 10^{19}  \rm ergs/sec/Hz/\msun ~.
\label{planck_alpha}
\end{eqnarray}

\noindent
This value for $\alpha_{850\mu \rm m}$ obtained from the Planck data in Taurus is remarkably similar to that found above (Equation \ref{alpha}) in the samples of nearby star forming galaxies, ULIRGs and z $\sim 2$ SMGs. Using Planck data from the Galaxy, \cite{pla11a} found $\tau_{250\mu \rm m} / N_{\rm H} = 0.92 \pm 0.05 \times 10^{-25}$ cm$^{2}$ near the solar circle. This determination 
at low angular resolution and covering a large range of galactic latitude is strongly weighted toward the HI phase in the solar neighborhood. Hence it is not surprising that it 
agrees better with the value found in Taurus for the atomic gas.  

\subsection{Expected Submm Fluxes as a function of Redshift}

Combining Equations \ref{lnu_eqn} and A4, the expected flux density at observed frequency $\nu_{obs}$ is given by 

  \begin{eqnarray}
  S_{\nu_{obs}} &=& 0.563 ~ {M_{\rm mol} \over 10^{10}\msun} ~ (1+z)^{4.8} ~   \left({\nu_{obs} \over{ \nu_{850\mu \rm m}}}\right)^{3.8}   (d_{L} [\rm Gpc])^{-2} \nonumber \\
 && \times ~\left\{{\alpha_{850} \over{6.7\times10^{19} }} \right\}    ~{\it{\Gamma_{RJ}} \over{\it{\Gamma_{0}}}}~   ~\rm mJy \\
\rm for  && \lambda_{rest} \gtrsim 250  ~\mu \rm m  \nonumber .
\label{snu_alpha}
  \end{eqnarray}

\noindent We note that the empirical calibration of $\alpha_{850}$ was obtained from z $\simeq 0$ galaxies which have a non-negligible 
RJ departure ($\Gamma_{0} \sim 0.7$) which must be normalized out  (i.e. the y-axis intercept in Figure \ref{rj}). This is the term $\Gamma_0 = \Gamma_{RJ} (0,T_d,\nu_{850})$ in the equation above. 

The restriction $ \lambda_{rest} \gtrsim 250  ~\mu \rm m$ is intended to ensure that one is on the RJ tail and that the dust is likely to be optically thin. If the dust is extremely 
cold one might need to be more restrictive and in the case of the most extreme ULIRGs the dust is probably optically thick 
to even longer wavelengths.  Analogous expressions are readily obtained for the other ALMA bands. 

Figure \ref{alma_obs} shows the expected flux as a function of redshift
for the ALMA bands at 100, 145, 240 and 350 GHz (Bands 3, 4,  6 and 7). At low z, the increasing luminosity 
distance leads to reduced flux as z increases. However, above z = 1 the well known ``negative k-correction" causes the flux per unit ISM mass to increase at higher
z as one moves up the far infrared SED towards the peak at $\lambda \sim 100 \mu$m. Figure \ref{alma_obs}
shows that the 350 GHz flux density plateaus at z = 1 and then decreases above z = 2. The latter is due to the fact that at higher redshift observed frame 350 GHz is approaching the rest frame far infrared peak (and no longer on the $\nu^2$ RJ tail). This is the factor $\Gamma_{RJ}$ coming in for 25 K dust.

 At  redshifts above 2.5, Figure \ref{alma_obs} indicates that one needs to shift to a lower frequency  
band, e.g.  240, 145 or 100 GHz, in order to avoid the large and uncertain $\Gamma_{RJ}$ corrections. Since future studies similar to that pursued here 
will push to higher redshifts, we have included the lower frequency bands in Figures \ref{rj} and \ref{alma_obs}.

Inverting Equation A13, the estimation of masses from observed flux densities can be done using 
  \begin{eqnarray}
M_{\rm mol}   &=& 1.78 ~S_{\nu_{obs}}[\rm mJy]  ~ (1+z)^{-4.8} ~   \left({ \nu_{850\mu \rm m}\over{\nu_{obs} }}\right)^{3.8}   ({\it{d}_L \rm{[Gpc ]}})^{2} \nonumber \\
 && \times ~\left\{{6.7\times10^{19}\over{\alpha_{850}  }} \right\}    ~{\it{\Gamma_{0}} \over{\it{\Gamma_{RJ}}}}~   ~10^{10}\msun  ~~ for~~  \lambda_{rest} > 250 \mu m  ~. \label{mass_eq}
  \end{eqnarray}
\noindent  The restriction $ \lambda_{rest} \gtrsim 250  ~\mu \rm m$ is intended to ensure that one stays on the Rayleigh-Jeans tail. 

In the above analysis we used a single (standard) Galactic conversion factor $\alpha_{CO}$ to convert observed CO(1-0) luminosity to gas mass. As discussed in 
\cite{sol05}, low-z studies of ULIRGs have led to the suggestion that the conversion factor could be several times smaller \citep{dow93,bry99}. This can arise in the ULIRGs 
if the gas is concentrated in the nuclear regions (as a result of dissipative galaxy merging) and the molecular emission linewidths can be broadened
by the galactic dynamics associated with the stellar mass -- not just the self-gravitating gas mass as in individual GMCs in which the standard conversion 
factor was derived. In addition, the mean gas temperature and density ($\rho$) may be different in the ULIRG nuclei as a result of the intense 
star formation activity, and the $\alpha_{CO}$ should vary as $<\rho>^{1/2}/T_k$ \citep[][Equation 8.5]{dic86,sco12}. 

Given the results obtained here which clearly 
show a quite similar $\alpha_{850\mu \rm m}$ in all three samples of galaxies, it would appear that there is little basis for using different $\alpha_{CO}$ for 
normal and star bursting galaxies -- at least when considering global measurements. 
For the high z SMGs, it is not obvious that 
the lower $\alpha_{CO}$ (often used in low-z ULIRGs) is appropriate since it is uncertain that the bulk of the molecular gas in the SMGs is similarly concentrated. 
Our restriction to CO(1-0) in the above sample was specifically intended to avoid sensitivity to the presence of high excitation 
gas, and to sample the larger, presumably extended masses of cold gas. Indeed, the ratio of dust emission to 
gas mass is similar to that obtained in low z galaxies. 

\subsection{Summary -- an approximately constant RJ mass-to-light ratio}

In the preceding sections, we have presented the physical explanation and, more importantly, strong 
empirical justification for using the long wavelength RJ dust emission in galaxies as a linear 
probe of ISM mass. The most substantial determination of the dust RJ spectral slope is that 
obtained by Planck from observations of the Galaxy \citep{pla11a,pla11b}, indicating a dust emissivity index 
$\beta = 1.8 \pm 0.1$ with no strong evidence of variation in Galactic radius or between atomic and molecular regions. 
Secondly, both the Planck data and measurements for nearby local galaxies, including both normal star forming and star bursting systems, 
indicate a similar constant of proportionality $\alpha_{850}$ for the dust emission at rest frame 850$\mu$m per unit mass of ISM. Lastly, 
we find that for a large sample of SMGs at z = 1.4 to 3, their ratio of rest frame 850$\mu$m per unit mass of ISM is essentially identical to that obtained 
for local galaxies. The similarities of these values of $\alpha_{850}$ argue strongly that for global ISM masses: the dust emissivity at long wavelengths, 
the dust-to-gas mass ratio and the mass-weighted dust temperatures vary little. 

The submm flux to dust mass ratio is expected to vary linearly with dust temperature. In practice, the overall range of $T_d$ for the bulk of the 
mass of ISM is very small, since it requires very large increases in radiative heating to increase the dust temperatures ($T_d$ varies approximately 
as the 1/5 - 1/6 power of the radiation energy density). As noted above, the extensive surveys of local galaxies using Herschel find a 
range of $T_d \sim 15 - 30$ K \citep{dun11,dal12,aul13}. Where we have needed to specify a dust temperature (e.g. for the R-J correction)
we have adopted 25 K, so we expect the uncertainties in the derived masses averaged on galaxy scales will be less than $\sim 25$\%. 

These calibrations include normal to star bursting systems and low to moderate redshift; they lay a solid foundation for using measurements of the RJ dust emission to probe galactic ISM masses. 
ALMA enables this technique for high redshift surveys, providing high sensitivity and the requisite angular resolution to avoid source confusion.

\subsection{Cautions}

It is important to recognize that even for those objects detected in SPIRE, the SPIRE data can not be used to reliably estimate ISM masses (along the 
lines as done here) for the z = 1 and 2 samples. For those redshifts, the SPIRE data will be probing near the rest frame far infrared luminosity peak -- {\it not safely on the RJ tail and not necessarily 
optically thin}.
The longest wavelength channel (500$\mu$m or 600 GHz) will be probing rest frame 170$\mu$m for z = 2; for such measurements, so there will be substantial uncertainty in the mass estimate, depending on the assumed value of the dust temperature (see Figure \ref{rj}). (In addition, the 500$\mu$m SPIRE data has relatively high source confusion on account of the large beam size.) 

Often, the far infrared SEDs are analyzed by fitting either modified black body curves or libraries of dust SEDs to the observed SEDs \cite[e.g.][]{dra07b,dac11,mag12,mag12a}. 
In essentially all instances the intrinsic SEDs used for fitting are taken to be optical thin. They thus do not include the attenuation expected near the far infrared 
peak associated with optically thick dust, instead attributing the drop at short wavelengths to a lack of high temperature grains. The $T_D$ determined in these 
cases is not even a luminosity-weighted $T_D$ of all the dust but just the dust above $\tau \sim 1$. 

\emph{Fitting the observed spectral energy distribution (SED) to derive an effective dust temperature is {\bf not} a reliable approach} -- near the far infrared peak, the temperature characterizing the emission is 'luminosity-weighted' (i.e. grains undergoing strong radiative heating) rather than mass-weighted. Hence, the derived $T_d$ will not reflect the temperature appropriate to the bulk of the ISM mass. Or, put another way, the flux measured near the peak is simply a measure of luminosity -- not mass. At high redshifts, the
large SPIRE beam at $500\mu$m results in severe source confusion at the expected flux levels; hence reliable flux measurements for individual galaxies are difficult. At z $> 2$ ALMA resolution and sensitivity 
are required and one must observe at $\nu \leq 350$ GHz to be on the RJ tail of the dust emission. 
 
One might be concerned that some of the correlation between the SPIRE 500$\mu$m continuum and the CO(1-0) in our local galaxy samples was due to emission lines contributing 
substantially to the continuum flux in the SPIRE data. However, scaling the CO(1-0) fluxes (given in the above tables) to the frequencies 
covered in the 500$\mu$m filter (having width $\lambda / \Delta \lambda$  = 2.5) indicates that the CO lines will contribute less than $10^{-3}$ of the total continuum flux. 
At rest wavelengths longer than 2mm, the line contamination becomes an issue since the line fluxes 
decrease less rapidly than $\lambda ^{-2}$ while the dust continuum decreases as $\sim \lambda ^{-3.8}$. In fact, the Planck imaging of nearby molecular clouds 
show positive excess residuals at the level of 10\% relative to the dust emission in the 100 GHz band \citep[][attributed to CO(1-0) emission]{pla11b}. 

Lastly, we reiterate the caution that the calibration samples are intentionally restricted to objects with high 
stellar mass ($M_{stellar} > 5\times10^{10}$\msun); thus we are not probing lower metallicity systems where the dust-to-gas abundance ratio is likely to drop 
or where there could be significant molecular gas without CO \citep[see][]{bol13}. A similar calibration at lower metallicities will be more difficult given the lower CO 
and continuum fluxes and in fact, at very low metallicities it is quite likely that this technique will not be so robust. 

\section{Individual Galaxies and their Fluxes}\label{source_app}

In Tables \ref{lowz} - \ref{highz} we list the individual flux measurements, and galaxy properties are summarized for all 145 
galaxies in our survey. The objects are taken from the COSMOS survey field \citep{sco_ove} and the galaxy properties are from the 
latest photometric redshift catalog \citep{ilb13}. This catalog has high accuracy photometric redshifts based on very deep 34 band photometry, including near infrared photometry from the Ultra-Vista survey. 
See \cite{ilb13} and \cite{lai15} for discussion of the accuracy of the redshifts and the stellar masses of the  galaxies. The SFRs in Column 9 are from \cite{lee15} where there are two band Herschel 
detections and from \cite{lai15} using the 
UV continuum and optical/UV continuum SEDs. 

The galaxy ID \# given in column 1 is taken from the most up to date COSMOS photometric redshift catalog \citep{lai15}. Columns 5 and 7 in each table list integrated and peak flux measurements for apertures of up to 2.5\arcsec (Low-z and Mid-z) and 2\arcsec (High-z) radius centered on the galaxy position. The aperture sizes are intended to include most of a galactic disk ($\sim 10$ kpc). The noise estimate in both cases is from the measured dispersion in the integrated and peak flux measurements obtained 
for 100 displaced off-center apertures of the same size in each individual image. The signal-to-noise ratio given in Column 7 is the better of those obtained
from the integrated or peak flux measurement; it is the ratio of the signal in Columns 5 and 7 to the measured noise given in Columns 6 and 8. Columns 10 - 12 give the 
galaxy stellar mass, SFR and sSFR (relative to the Main Sequence at the same redshift and $M_{stellar}$ with the Main Sequence taken from \cite{lee15}. In the last column, 
the derived ISM molecular mass is given. Limits on the masses are at 2$\sigma$ and 3.6$\sigma$,
depending on whether the better SNR (column 9) was obtained for the integrated or peak flux measurement. The detection thresholds of 2 and 3.6 $\sigma$ are chosen such that the chance of a spurious detection across the entirety of each sample is less than $\sim10$\% (based on the measured noise in each individual image).

\begin{deluxetable*}{rrrrrrrrrcccr} 
%\tabletypesize{0.2}
%\normalsize
%\rotate
\tablecaption{{\bf Low-z Galaxy Sample in Band 7}}
\tablewidth{0pt}
\tablehead{ 
 \colhead{\#} &  \colhead{RA (2000)} & \colhead{Dec (2000)} & \colhead{z\tablenotemark{a} } &\colhead{S$_{tot}$} &\colhead{$\sigma_{tot}$} & \colhead{S$_{peak}$} &\colhead{$\sigma_{pix}$} & \colhead{SNR\tablenotemark{b}}  & \colhead{M$_{*}$\tablenotemark{a}} &  \colhead{Log(SFR)\tablenotemark{a}} & \colhead{sSFR} &  \colhead{M$_{\rm mol}$}  \\
 \colhead{} & \colhead{\deg} & \colhead{\deg} & \colhead{}  & \colhead{mJy} & \colhead{mJy}  & \colhead{mJy} & \colhead{mJy}  & \colhead{} &  \colhead{ $10^{11}$\msun} &  \colhead{\msun yr$^{-1}$}& \colhead{/sSFR$_{MS}$} & \colhead{$10^{10}$\msun} }
\startdata 
  890088 &   149.8157 &     2.6503 &      1.02 &      0.38 &      0.19 &      0.71 &      0.17 &      4.15 &      0.28 &      1.19 &      0.65 &        3.22$\pm$0.78 \\
  846385 &   149.5421 &     2.5816 &      1.01 &      0.37 &      0.18 &      0.62 &      0.18 &      2.00 &      0.39 &      1.43 &      1.00 &             $<$ 2.41 \\
  405136 &   150.7642 &     1.9077 &      1.02 &      0.26 &      0.13 &      0.33 &      0.10 &      2.00 &      0.28 &      1.35 &      0.92 &             $<$ 1.40 \\
  471515 &   149.8180 &     2.0132 &      1.04 &      0.24 &      0.12 &      0.37 &      0.11 &      2.00 &      0.40 &      1.40 &      0.88 &             $<$ 1.46 \\
  257225 &   149.9700 &     1.6731 &      1.01 &      0.20 &      0.10 &      0.29 &      0.10 &      2.00 &      0.32 &      1.11 &      0.51 &             $<$ 1.39 \\
  354566 &   149.9792 &     1.8278 &      1.19 &      0.83 &      0.32 &      0.36 &      0.11 &      2.55 &      0.38 &      1.47 &      0.85 &        3.99$\pm$1.57 \\

  737603 &   150.0182 &     2.4177 &      1.12 &      0.31 &      0.16 &      0.51 &      0.15 &      2.00 &      0.52 &      1.20 &      0.46 &             $<$ 2.08 \\
  787724 &   149.8601 &     2.4920 &      1.19 &      0.25 &      0.12 &      0.58 &      0.15 &      3.80 &      0.52 &      1.48 &      0.78 &        2.72$\pm$0.72 \\
  543756 &   150.0946 &     2.1280 &      1.22 &      0.28 &      0.14 &      0.62 &      0.14 &      4.28 &      0.60 &      1.44 &      0.66 &        2.99$\pm$0.70 \\
  805480 &   150.0256 &     2.5211 &      1.03 &      0.30 &      0.15 &      0.46 &      0.16 &      2.00 &      0.79 &      1.51 &      0.96 &             $<$ 2.24 \\

  687914 &   149.7395 &     2.3413 &      1.01 &      0.33 &      0.17 &      0.63 &      0.16 &      3.96 &      0.38 &      1.47 &      1.10 &        2.86$\pm$0.72 \\
  711951 &   149.8388 &     2.3775 &      1.21 &      0.30 &      0.15 &      2.63 &      0.16 &     16.47 &      0.39 &      1.84 &      1.90 &       12.78$\pm$0.78 \\
  800281 &   150.0747 &     2.5130 &      1.26 &      0.31 &      0.16 &      0.47 &      0.15 &      2.00 &      0.39 &      1.78 &      1.60 &             $<$ 2.18 \\
  873708 &   150.1164 &     2.6253 &      1.02 &      0.29 &      0.14 &      0.57 &      0.18 &      2.00 &      0.39 &      1.63 &      1.54 &             $<$ 2.38 \\

  599889 &   150.3524 &     2.2101 &      1.23 &      2.48 &      0.50 &      0.45 &      0.15 &      5.02 &      0.99 &      2.03 &      2.28 &       12.10$\pm$2.41 \\
  671444 &   150.2712 &     2.3184 &      1.16 &      1.21 &      0.48 &      0.63 &      0.15 &      2.53 &      0.85 &      1.88 &      1.84 &        5.80$\pm$2.29 \\
  918541 &   149.8367 &     2.6918 &      1.01 &      1.07 &      0.46 &      0.64 &      0.18 &      2.35 &      0.54 &      1.60 &      1.32 &        4.82$\pm$2.05 \\
  984597 &   150.1099 &     2.7994 &      1.09 &      0.38 &      0.19 &      0.50 &      0.18 &      2.00 &      0.44 &      1.64 &      1.38 &             $<$ 2.53 \\
  802829 &   149.7721 &     2.5171 &      1.14 &      2.62 &      0.40 &      0.61 &      0.17 &      6.60 &      0.72 &      1.75 &      1.45 &       12.47$\pm$1.89 \\
  806656 &   150.2180 &     2.5217 &      1.20 &      2.00 &      0.41 &      0.60 &      0.16 &      4.91 &      0.74 &      1.88 &      1.77 &        9.66$\pm$1.97 \\
  842212 &   150.0628 &     2.5741 &      1.01 &      0.44 &      0.17 &      0.66 &      0.17 &      3.98 &      0.95 &      1.80 &      1.84 &        1.99$\pm$0.50 \\
  483264 &   150.2705 &     2.0338 &      1.20 &      0.27 &      0.13 &      0.32 &      0.11 &      2.00 &      0.41 &      1.77 &      1.63 &             $<$ 1.66 \\

  541203 &   149.6961 &     2.1223 &      1.37 &      3.02 &      0.56 &      0.55 &      0.15 &      5.40 &      1.28 &      2.11 &      2.21 &       15.18$\pm$2.81 \\
  914622 &   150.0537 &     2.6861 &      1.03 &      0.37 &      0.18 &      0.57 &      0.17 &      2.00 &      2.88 &      1.94 &      2.21 &             $<$ 2.37 \\
  983395 &   150.6615 &     2.7970 &      1.18 &      1.26 &      0.63 &      0.73 &      0.18 &      2.00 &      1.05 &      1.95 &      2.00 &        6.04$\pm$3.02 \\
  985857 &   150.3477 &     2.8015 &      1.25 &      1.10 &      0.49 &      0.54 &      0.17 &      3.15 &      1.39 &      1.91 &      1.60 &        5.37$\pm$1.70 \\
  422455 &   149.8490 &     1.9337 &      1.24 &      0.74 &      0.24 &      0.28 &      0.11 &      3.10 &      2.12 &      1.87 &      1.44 &        3.60$\pm$1.16 \\
  434350 &   150.3446 &     1.9540 &      1.24 &      0.86 &      0.41 &      0.34 &      0.11 &      2.11 &      1.86 &      1.77 &      1.15 &        4.22$\pm$2.00 \\
  505822 &   149.6633 &     2.0667 &      1.24 &      1.75 &      0.39 &      0.43 &      0.10 &      4.51 &      1.66 &      2.09 &      2.43 &        8.58$\pm$1.90 \\
  340442 &   150.1541 &     1.8042 &      1.11 &      2.28 &      1.13 &      1.04 &      0.11 &      2.01 &      1.59 &      1.96 &      2.15 &       10.70$\pm$5.32 \\

  868539 &   149.8634 &     2.6178 &      1.32 &      3.87 &      0.79 &      1.81 &      0.17 &      4.89 &      0.35 &      2.04 &      2.74 &       19.31$\pm$3.95 \\
  288391 &   150.4103 &     2.6264 &      1.15 &      0.32 &      0.16 &      2.24 &      0.18 &     12.61 &      0.28 &      1.96 &      3.14 &       10.68$\pm$0.85 \\
  960580 &   149.9191 &     2.7605 &      1.16 &      2.66 &      0.50 &      0.58 &      0.17 &      5.36 &      0.35 &      2.07 &      3.67 &       12.69$\pm$2.37 \\
  409265 &   149.7843 &     1.9128 &      1.16 &      0.69 &      0.27 &      0.33 &      0.11 &      2.60 &      0.39 &      2.11 &      3.84 &        3.31$\pm$1.27 \\
  351159 &   150.0908 &     1.8211 &      1.00 &      0.20 &      0.10 &      0.35 &      0.10 &      2.00 &      0.35 &      1.91 &      3.14 &             $<$ 1.39 \\

  601886 &   149.8117 &     2.2123 &      1.22 &      0.30 &      0.15 &      0.50 &      0.16 &      2.00 &      0.95 &      2.35 &      4.86 &             $<$ 2.32 \\
  237348 &   149.4016 &     2.4509 &      1.19 &      3.08 &      0.50 &      1.85 &      0.16 &      6.18 &      0.76 &      2.24 &      4.09 &       14.89$\pm$2.41 \\
  781580 &   149.9086 &     2.4833 &      1.25 &      0.52 &      0.15 &      0.56 &      0.15 &      3.40 &      0.75 &      2.12 &      2.91 &        2.53$\pm$0.74 \\
  134318 &   150.2174 &     2.1141 &      1.13 &      0.90 &      0.45 &      0.80 &      0.15 &      2.00 &      0.44 &      2.22 &      4.90 &        4.24$\pm$2.12 \\
  560724 &   150.1236 &     2.1498 &      1.17 &      1.83 &      0.42 &      0.41 &      0.15 &      4.37 &      0.83 &      2.15 &      3.40 &        8.78$\pm$2.01 \\
  585275 &   150.3987 &     2.1885 &      1.04 &      2.98 &      0.39 &      2.02 &      0.15 &      7.67 &      0.78 &      2.18 &      4.43 &       13.56$\pm$1.77 \\
  969105 &   150.5307 &     2.7755 &      1.36 &      0.89 &      0.35 &      0.65 &      0.18 &      2.55 &      0.96 &      2.33 &      3.83 &        4.47$\pm$1.75 \\
  831023 &   150.4179 &     2.5585 &      1.21 &      2.55 &      0.98 &      0.54 &      0.16 &      3.25 &      0.71 &      2.28 &      4.46 &       12.35$\pm$3.80 \\
    6496 &   149.9741 &     1.6435 &      1.04 &      1.23 &      0.40 &      0.47 &      0.10 &      3.05 &      0.76 &      2.16 &      4.24 &        5.60$\pm$1.84 \\
  418048 &   150.3503 &     1.9282 &      1.20 &      2.14 &      0.44 &      0.46 &      0.11 &      4.81 &      0.83 &      2.15 &      3.21 &       10.35$\pm$2.15 \\
  108065 &   150.0087 &     2.0257 &      1.19 &      1.45 &      0.65 &      0.35 &      0.10 &      2.23 &      0.76 &      2.09 &      2.93 &        7.02$\pm$3.14 \\
  269311 &   149.4324 &     1.6924 &      1.26 &      2.86 &      0.54 &      1.30 &      0.11 &      5.32 &      0.61 &      2.16 &      3.24 &       14.08$\pm$2.65 \\

  160476 &   149.5986 &     2.2004 &      1.20 &      3.94 &      0.95 &      1.05 &      0.15 &      4.17 &      2.24 &      2.44 &      5.65 &       19.06$\pm$4.57 \\
  872762 &   150.1296 &     2.6214 &      1.41 &      3.57 &      1.46 &      0.96 &      0.16 &      2.44 &      2.31 &      2.28 &      2.91 &       18.11$\pm$7.42 \\
  916658 &   150.2231 &     2.6903 &      1.30 &      0.39 &      0.19 &      0.52 &      0.17 &      2.00 &      1.24 &      2.42 &      5.01 &             $<$ 2.60 \\
  811432 &   150.0425 &     2.5266 &      1.18 &      2.56 &      0.36 &      0.77 &      0.16 &      4.72 &      1.85 &      2.13 &      2.85 &       12.30$\pm$2.61 \\
  485345 &   150.1895 &     2.0370 &      1.18 &      2.31 &      0.35 &      0.75 &      0.11 &      6.52 &      1.12 &      2.19 &      3.44 &       11.11$\pm$1.70 \\
  370733 &   150.4130 &     1.8517 &      1.20 &      1.72 &      0.21 &      0.35 &      0.11 &      8.33 &      1.26 &      2.11 &      2.79 &        8.32$\pm$1.00 \\

  627524 &   149.9813 &     2.2536 &      1.36 &      0.27 &      0.14 &      1.28 &      0.15 &      8.38 &      0.22 &      2.36 &      6.77 &        6.42$\pm$0.77 \\
  344653 &   150.5036 &     1.8118 &      1.16 &      1.95 &      0.27 &      1.16 &      0.11 &      7.11 &      0.36 &      2.34 &      6.58 &        9.35$\pm$1.31 \\

  504172 &   149.8325 &     2.0660 &      1.15 &      2.84 &      0.28 &      1.72 &      0.15 &     10.22 &      0.66 &      2.44 &      7.06 &       13.51$\pm$1.32 \\
    9254 &   150.0120 &     1.6521 &      1.30 &      1.44 &      0.29 &      0.69 &      0.11 &      6.38 &      0.42 &      2.65 &     10.92 &        7.16$\pm$1.12 \\

  570293 &   150.0981 &     2.1658 &      1.20 &      2.11 &      0.50 &      1.46 &      0.15 &      9.64 &      2.06 &      2.48 &      6.15 &       10.19$\pm$1.06 \\

   \\ 
\enddata \label{lowz} 
\tablenotetext{a}{The photometric redshifts and stellar masses of the galaxies are from \cite{ilb13} and the accuracy of those quantities is discussed in detail there. The SFRs are derived from the rest frame UV continuum and the infrared using Herschel PACS and SPIRE data as detailed in \cite{sco13}. All of the galaxies have greater than 10$\sigma$ photometry measurements so the uncertainties in M$_*$ and SFR associated with their measurements are always less than 10\%. As discussed in \cite{ilb13} the uncertainties in models used to 
derive the M$_*$ and SFR from the photometry are larger but generally less than a factor 2.}
\tablenotetext{b}{SNR$_{tot}$ and SNR$_{peak}$ are calculated separately and the column SNR lists the larger in absolute magnitude of those two SNRs. Note that we let the SNR be negative in cases where the flux estimate is negative so that several sigma negative flux values don't end up with a positive SNR above the detection thresholds.  }
\end{deluxetable*}

\begin{deluxetable*}{rrrrrrrrrcccr}
%\tabletypesize{0.2}
%\normaltsize
%\rotate
\tablecaption{\bf{Mid-z Galaxy Sample in Band 7}  }
%\tablewidth{0pt}

\tablehead{ 
 \colhead{\#} &  \colhead{RA (2000)} & \colhead{Dec (2000)} & \colhead{z\tablenotemark{a} } &\colhead{S$_{tot}$} &\colhead{$\sigma_{tot}$} & \colhead{S$_{peak}$} &\colhead{$\sigma_{pix}$} & \colhead{SNR\tablenotemark{b}}  & \colhead{M$_{*}$\tablenotemark{a}} &  \colhead{Log(SFR)\tablenotemark{a}} & \colhead{sSFR} &   \colhead{M$_{\rm mol}$}  \\
 \colhead{} & \colhead{\deg} & \colhead{\deg} & \colhead{}  & \colhead{mJy} & \colhead{mJy}  & \colhead{mJy} & \colhead{mJy}  & \colhead{} &  \colhead{ $10^{11}$\msun} &  \colhead{\msun yr$^{-1}$}& \colhead{/sSFR$_{MS}$} &\colhead{$10^{10}$\msun} }
\startdata 

  399465 &   150.4691 &     1.8996 &      2.20 &      0.26 &      0.13 &      0.34 &      0.14 &      2.00 &      0.35 &      1.86 &      0.69 &             $<$ 2.29 \\
  479473 &   150.0251 &     2.0288 &      2.17 &      0.24 &      0.12 &      0.42 &      0.14 &      2.00 &      0.26 &      1.70 &      0.56 &             $<$ 2.25 \\
   35012 &   149.9635 &     1.7615 &      2.00 &      0.40 &      0.15 &      0.39 &      0.15 &      2.55 &      0.32 &      1.81 &      0.78 &        2.18$\pm$0.86 \\
  829041 &   149.7756 &     2.5571 &      2.03 &      1.10 &      0.24 &      0.36 &      0.12 &      4.58 &      0.27 &      1.82 &      0.83 &        5.97$\pm$1.30 \\
  612589 &   149.9561 &     2.2301 &      2.43 &      0.26 &      0.13 &      0.30 &      0.12 &      2.00 &      0.21 &      1.88 &      0.73 &             $<$ 1.97 \\
  708203 &   150.7118 &     2.3724 &      2.28 &      0.26 &      0.13 &      0.39 &      0.12 &      2.00 &      0.31 &      1.72 &      0.49 &             $<$ 1.93 \\

  306429 &   150.1298 &     1.7536 &      2.31 &      1.13 &      0.52 &      0.60 &      0.15 &      4.11 &      0.50 &      1.93 &      0.66 &        6.23$\pm$1.52 \\
  777598 &   149.9130 &     2.4806 &      2.89 &      0.48 &      0.20 &      0.25 &      0.06 &      2.43 &      0.72 &      2.09 &      0.72 &        6.55$\pm$2.70 \\
  524944 &   150.0317 &     2.0987 &      2.34 &      1.63 &      0.49 &      0.59 &      0.12 &      3.32 &      0.82 &      2.01 &      0.68 &        8.97$\pm$2.70 \\

  348260 &   150.2028 &     1.8191 &      2.39 &      1.74 &      0.37 &      0.65 &      0.14 &      4.55 &      1.59 &      2.22 &      0.96 &        9.63$\pm$2.11 \\
  715833 &   150.5795 &     2.3850 &      2.71 &      0.46 &      0.20 &      0.29 &      0.06 &      2.29 &      1.15 &      2.04 &      0.59 &        6.44$\pm$2.81 \\

  323041 &   149.8165 &     1.7798 &      2.11 &      1.54 &      0.47 &      0.73 &      0.14 &      5.18 &      0.33 &      2.29 &      2.08 &        8.37$\pm$1.62 \\
  961356 &   149.6007 &     2.7629 &      2.37 &      0.27 &      0.14 &      0.42 &      0.13 &      2.00 &      0.16 &      2.29 &      2.34 &             $<$ 2.09 \\
  374178 &   149.7167 &     1.8609 &      2.28 &      0.89 &      0.12 &      0.77 &      0.07 &      7.10 &      0.32 &      2.34 &      2.01 &       13.04$\pm$1.84 \\
  608918 &   150.1193 &     2.2241 &      2.25 &      1.66 &      0.33 &      0.46 &      0.12 &      4.99 &      0.17 &      1.95 &      1.14 &        9.12$\pm$1.83 \\
  759305 &   150.3527 &     2.4511 &      1.91 &      0.26 &      0.13 &      0.35 &      0.12 &      2.00 &      0.24 &      2.09 &      1.82 &             $<$ 1.92 \\
  516419 &   149.7464 &     2.0845 &      2.14 &      0.22 &      0.11 &      0.36 &      0.12 &      2.00 &      0.38 &      2.10 &      1.25 &             $<$ 1.94 \\

  401783 &   149.9235 &     1.9038 &      2.16 &      3.08 &      0.50 &      1.55 &      0.14 &      6.16 &      0.87 &      2.40 &      1.94 &       16.80$\pm$2.73 \\
  414218 &   149.9098 &     1.9234 &      2.07 &      1.46 &      0.03 &      0.57 &      0.13 &     42.61 &      0.40 &      2.01 &      1.04 &        7.91$\pm$0.19 \\
  444936 &   149.7446 &     1.9724 &      2.30 &      1.41 &      0.44 &      0.66 &      0.15 &      3.23 &      0.78 &      2.30 &      1.37 &        7.76$\pm$2.41 \\
   95500 &   149.8838 &     1.9812 &      2.13 &      2.31 &      0.60 &      0.74 &      0.15 &      3.83 &      0.79 &      2.25 &      1.41 &       12.60$\pm$3.29 \\
  254938 &   150.3699 &     1.6707 &      2.01 &      0.77 &      0.24 &      0.46 &      0.13 &      3.43 &      0.79 &      2.09 &      1.11 &        4.16$\pm$1.21 \\
  818426 &   150.7220 &     2.5419 &      2.30 &      0.49 &      0.14 &      0.47 &      0.12 &      3.42 &      0.75 &      2.51 &      2.26 &        2.70$\pm$0.79 \\
  482039 &   150.1189 &     2.0321 &      2.15 &      0.23 &      0.12 &      0.39 &      0.12 &      2.00 &      0.84 &      2.11 &      1.00 &             $<$ 1.92 \\
  575173 &   149.9899 &     2.1741 &      2.01 &      1.03 &      0.50 &      0.46 &      0.12 &      2.05 &      0.54 &      2.28 &      1.91 &        5.58$\pm$2.72 \\
  672025 &   150.0164 &     2.3210 &      2.33 &      2.73 &      0.47 &      1.51 &      0.12 &      5.77 &      0.52 &      2.33 &      1.60 &       15.03$\pm$2.60 \\
  254150 &   150.0934 &     2.5073 &      2.22 &      3.74 &      0.49 &      2.21 &      0.12 &      7.58 &      0.76 &      2.21 &      1.21 &       20.48$\pm$2.70 \\
  514900 &   150.4552 &     2.0835 &      2.78 &      1.41 &      0.25 &      0.70 &      0.12 &      5.75 &      0.56 &      2.33 &      1.33 &        7.95$\pm$1.38 \\

  283400 &   149.6042 &     1.7164 &      2.05 &      4.18 &      0.02 &      2.95 &      0.15 &    174.99 &      2.37 &      2.41 &      1.96 &       22.67$\pm$0.13 \\
  311139 &   149.7768 &     1.7610 &      2.30 &      1.86 &      0.54 &      1.14 &      0.15 &      7.85 &      1.28 &      2.47 &      1.87 &       10.23$\pm$1.30 \\
  277716 &   149.4757 &     2.5882 &      2.04 &      2.14 &      0.78 &      0.93 &      0.13 &      2.73 &      2.63 &      2.36 &      1.72 &       11.58$\pm$4.24 \\
  909889 &   149.6367 &     2.6824 &      2.23 &      0.43 &      0.16 &      0.44 &      0.13 &      2.77 &      1.10 &      2.23 &      1.18 &        2.36$\pm$0.85 \\
  919588 &   150.1266 &     2.6961 &      2.22 &      4.62 &      1.22 &      3.25 &      0.13 &      3.79 &      2.32 &      2.45 &      1.82 &       25.26$\pm$6.66 \\
  932436 &   150.3178 &     2.7165 &      2.58 &      4.83 &      0.47 &      2.05 &      0.12 &     10.21 &      2.80 &      2.46 &      1.43 &       26.97$\pm$2.64 \\
  969701 &   149.5260 &     2.7769 &      2.09 &      1.42 &      0.30 &      0.64 &      0.12 &      4.75 &      1.29 &      2.38 &      1.82 &        7.70$\pm$1.62 \\
  830116 &   150.5398 &     2.5586 &      1.84 &      2.75 &      1.24 &      1.14 &      0.13 &      2.22 &      2.64 &      2.30 &      1.87 &       14.72$\pm$6.64 \\
  561437 &   150.5270 &     2.1541 &      2.73 &      4.66 &      0.34 &      1.50 &      0.12 &     13.84 &      1.64 &      2.62 &      2.18 &       26.23$\pm$1.90 \\

  421924 &   150.3745 &     1.9364 &      2.34 &      1.66 &      0.55 &      0.90 &      0.15 &      6.01 &      0.31 &      2.77 &      5.19 &        9.17$\pm$1.52 \\

  464593 &   150.3526 &     2.0048 &      2.21 &      2.47 &      0.58 &      2.15 &      0.14 &      4.23 &      0.65 &      2.70 &      3.94 &       13.51$\pm$3.19 \\
  287250 &   149.6534 &     1.7231 &      2.84 &      3.07 &      0.60 &      2.20 &      0.15 &     14.97 &      0.71 &      2.66 &      2.67 &       17.40$\pm$1.16 \\
  903144 &   150.4592 &     2.6714 &      2.04 &      1.21 &      0.41 &      0.59 &      0.13 &      2.98 &      0.54 &      2.54 &      3.31 &        6.56$\pm$2.20 \\
  917423 &   149.9921 &     2.6934 &      2.12 &      2.98 &      1.20 &      1.44 &      0.13 &      2.49 &      0.72 &      2.82 &      5.47 &       16.20$\pm$6.51 \\
  953800 &   150.1102 &     2.7516 &      2.30 &      8.66 &      0.65 &      4.05 &      0.13 &     13.35 &      1.00 &      2.75 &      3.75 &       47.57$\pm$3.56 \\
  821753 &   150.2927 &     2.5466 &      2.60 &      1.31 &      0.52 &      0.61 &      0.12 &      2.53 &      0.57 &      2.67 &      2.89 &        7.31$\pm$2.90 \\
  274938 &   149.9980 &     2.5782 &      2.35 &      1.89 &      0.34 &      1.17 &      0.15 &      5.59 &      0.72 &      2.94 &      5.85 &       10.43$\pm$1.87 \\
  476581 &   150.3899 &     2.0247 &      2.73 &      1.53 &      0.30 &      1.31 &      0.06 &      5.08 &      0.86 &      2.95 &      5.05 &       21.37$\pm$4.20 \\
  562990 &   150.1608 &     2.1547 &      2.30 &      2.08 &      0.33 &      0.45 &      0.12 &      6.40 &      0.49 &      2.56 &      2.81 &       11.44$\pm$1.79 \\
  122443 &   149.6588 &     2.0720 &      2.29 &      5.42 &      0.56 &      2.36 &      0.12 &      9.62 &      0.72 &      2.68 &      3.34 &       29.78$\pm$3.09 \\
  514719 &   149.5358 &     2.0825 &      2.17 &      0.25 &      0.13 &      0.40 &      0.12 &      2.00 &      0.84 &      2.54 &      2.65 &             $<$ 1.97 \\

  338500 &   150.2649 &     1.8029 &      2.36 &      9.17 &      0.67 &      6.11 &      0.16 &     13.79 &      1.69 &      2.68 &      2.78 &       50.54$\pm$3.67 \\
  372039 &   150.4384 &     1.8561 &      2.58 &      9.46 &      0.61 &      7.01 &      0.15 &     15.60 &      1.23 &      2.82 &      3.54 &       52.75$\pm$3.38 \\
  427827 &   150.3416 &     1.9456 &      2.78 &      3.47 &      0.29 &      2.67 &      0.18 &     15.08 &      3.32 &      2.73 &      2.65 &       19.58$\pm$1.30 \\
  842737 &   150.6338 &     2.5783 &      2.67 &      8.24 &      0.73 &      5.29 &      0.12 &     11.23 &      1.32 &      2.87 &      3.92 &       46.21$\pm$4.11 \\
  932331 &   149.6556 &     2.7162 &      2.11 &      7.13 &      1.08 &      3.47 &      0.13 &      6.63 &      2.83 &      2.58 &      2.66 &       38.77$\pm$5.85 \\
  942076 &   150.1471 &     2.7315 &      2.42 &     14.33 &      0.39 &      5.73 &      0.13 &     36.79 &      1.44 &      2.82 &      3.70 &       79.24$\pm$2.15 \\
  264030 &   150.7057 &     2.5404 &      2.15 &      4.34 &      0.69 &      2.19 &      0.12 &      6.26 &      1.95 &      2.72 &      3.68 &       23.64$\pm$3.77 \\
  495704 &   149.9889 &     2.0533 &      1.92 &      5.90 &      0.61 &      2.06 &      0.12 &      9.61 &      2.04 &      2.76 &      4.98 &       31.73$\pm$3.30 \\
  723263 &   149.8893 &     2.3964 &      2.18 &      5.26 &      0.61 &      2.69 &      0.13 &      8.56 &      1.06 &      2.59 &      2.85 &       28.71$\pm$3.36 \\
  126711 &   149.6679 &     2.0874 &      2.30 &     11.42 &      0.59 &      4.96 &      0.13 &     19.47 &      1.74 &      2.91 &      4.94 &       62.76$\pm$3.22 \\
  518250 &   150.1799 &     2.0886 &      2.32 &      5.40 &      0.31 &      3.00 &      0.13 &     17.62 &      2.65 &      2.67 &      2.71 &       29.71$\pm$1.69 \\
  135052 &   150.4957 &     2.1162 &      2.21 &      4.89 &      0.67 &      2.30 &      0.12 &      7.29 &      1.23 &      2.70 &      3.44 &       26.73$\pm$3.66 \\

  408649 &   149.6658 &     1.9139 &      1.93 &      3.33 &      0.49 &      2.51 &      0.14 &     18.62 &      1.46 &      2.85 &      6.26 &       17.91$\pm$0.96 \\
  980250 &   150.0161 &     2.7924 &      1.80 &      7.13 &      1.20 &      2.99 &      0.13 &      5.96 &      1.76 &      2.84 &      6.98 &       37.96$\pm$6.37 \\
  815012 &   150.6034 &     2.5366 &      2.10 &      5.38 &      0.74 &      3.55 &      0.13 &      7.32 &      1.57 &      3.10 &      9.36 &       29.24$\pm$4.00 \\

  \\ 
\enddata\label{midz}
\tablenotetext{a}{The photometric redshifts and stellar masses of the galaxies are from \cite{ilb13} and the accuracy of those quantities is discussed in detail there. The SFRs are derived from the rest frame UV continuum and the infrared using Herschel PACS and SPIRE data as detailed in \cite{sco13}. All of the galaxies have greater than 10$\sigma$ photometry measurements so the uncertainties in M$_*$ and SFR associated with their measurements are always less than 10\%. As discussed in \cite{ilb13} the uncertainties in models used to 
derive the M$_*$ and SFR from the photometry are larger but generally less than a factor 2.}
\tablenotetext{b}{SNR$_{tot}$ and SNR$_{peak}$ are calculated separately and the column SNR lists the larger in absolute magnitude of those two SNRs. Note that we let the SNR be negative in cases where the flux estimate is negative so that several sigma negative flux values don't end up with a positive SNR above the detection thresholds.  }
\end{deluxetable*}

\begin{deluxetable*}{rrrrrrrrrcccr}
%\tabletypesize{0.2}
%\normaltsize
%\rotate
\tablecaption{\bf{High-z Galaxy Sample in Band 6}} 
\tablewidth{0pt}

\tablehead{ 
 \colhead{\#} &  \colhead{RA (2000)} & \colhead{Dec (2000)} & \colhead{z \tablenotemark{a}} &\colhead{S$_{tot}$} &\colhead{$\sigma_{tot}$} & \colhead{S$_{peak}$} &\colhead{$\sigma_{pix}$} & \colhead{SNR\tablenotemark{b}}  & \colhead{M$_{*}$\tablenotemark{a}} &  \colhead{Log(SFR)\tablenotemark{a}} &  \colhead{sSFR} &  \colhead{M$_{\rm mol}$}  \\
 \colhead{} & \colhead{\deg} & \colhead{\deg} & \colhead{}  & \colhead{mJy} & \colhead{mJy}  & \colhead{mJy} & \colhead{mJy}  & \colhead{} &  \colhead{ $10^{11}$\msun} &  \colhead{\msun yr$^{-1}$}&  \colhead{/sSFR$_{MS}$}& \colhead{$10^{10}$\msun} }
\startdata 

  566428 &   150.0300 &     2.1627 &      5.89 &      0.14 &      0.07 &      0.22 &      0.07 &      2.00 &      0.25 &      1.74 &      0.46 &             $<$ 2.51 \\
  457406 &   150.3921 &     1.9937 &      4.00 &      0.16 &      0.08 &      0.19 &      0.06 &      2.10 &      0.26 &      1.83 &      0.56 &        2.03$\pm$0.97 \\
  286380 &   150.0598 &     1.7217 &      4.35 &      0.12 &      0.06 &      0.31 &      0.07 &      4.69 &      0.31 &      1.94 &      0.67 &        3.77$\pm$0.80 \\

  477614 &   150.3071 &     2.0261 &      4.30 &      0.13 &      0.07 &      0.18 &      0.07 &      2.00 &      0.45 &      2.07 &      0.79 &             $<$ 2.41 \\
  249399 &   150.1373 &     2.4902 &      4.16 &      0.09 &      0.05 &      0.18 &      0.06 &      2.00 &      0.56 &      1.90 &      0.50 &             $<$ 2.16 \\

  735699 &   150.6181 &     2.4158 &      4.04 &      0.55 &      0.11 &      0.20 &      0.06 &      3.54 &      1.03 &      2.14 &      0.75 &        6.90$\pm$1.95 \\

  608706 &   150.5920 &     2.2251 &      4.85 &      0.11 &      0.05 &      0.21 &      0.07 &      2.00 &      0.16 &      2.35 &      2.37 &             $<$ 2.41 \\
  972851 &   149.9827 &     2.7821 &      4.82 &      0.10 &      0.05 &      0.14 &      0.06 &      2.00 &      0.24 &      2.30 &      1.72 &             $<$ 2.07 \\
  284164 &   150.5189 &     2.6097 &      4.21 &      0.13 &      0.07 &      0.14 &      0.06 &      2.00 &      0.33 &      2.48 &      2.27 &             $<$ 2.23 \\
  386988 &   150.1413 &     1.8805 &      4.71 &      0.30 &      0.11 &      0.17 &      0.06 &      2.89 &      0.18 &      2.04 &      1.08 &        3.71$\pm$1.28 \\
  256965 &   150.3371 &     1.6746 &      4.59 &      0.39 &      0.17 &      0.26 &      0.06 &      4.21 &      0.18 &      2.32 &      2.09 &        4.81$\pm$1.14 \\

   41128 &   149.3435 &     1.7836 &      5.59 &      0.13 &      0.06 &      0.16 &      0.06 &      2.00 &      0.45 &      2.55 &      2.35 &             $<$ 2.32 \\

  582526 &   149.8712 &     2.1871 &      4.55 &      0.12 &      0.06 &      0.21 &      0.06 &      2.00 &      3.66 &      2.48 &      1.48 &             $<$ 2.34 \\
  331108 &   150.4637 &     2.7859 &      4.49 &      1.98 &      0.35 &      1.12 &      0.06 &     19.54 &      1.02 &      2.53 &      1.85 &       24.34$\pm$1.25 \\

  901851 &   150.4011 &     2.6707 &      4.14 &      0.27 &      0.07 &      0.23 &      0.06 &      3.77 &      0.11 &      2.40 &      3.38 &        3.33$\pm$0.88 \\
  302769 &   150.0692 &     1.7477 &      4.33 &      0.58 &      0.15 &      0.27 &      0.06 &      3.88 &      0.12 &      2.46 &      3.65 &        7.12$\pm$1.84 \\
  307139 &   150.1546 &     1.7550 &      4.30 &      0.46 &      0.05 &      0.21 &      0.06 &      3.35 &      0.36 &      2.66 &      3.30 &        5.72$\pm$1.71 \\

  881017 &   150.4657 &     2.6361 &      3.54 &      2.24 &      0.40 &      1.33 &      0.06 &     23.27 &      0.50 &      2.90 &      5.10 &       28.95$\pm$1.24 \\

  468591 &   150.5352 &     2.0115 &      4.13 &      1.43 &      0.15 &      0.84 &      0.07 &      9.54 &      1.06 &      2.73 &      2.95 &       17.85$\pm$1.87 \\
  536066 &   150.4204 &     2.1177 &      3.96 &      1.34 &      0.11 &      0.99 &      0.07 &     11.91 &      2.48 &      2.89 &      3.86 &       16.87$\pm$1.42 \\
  564267 &   150.2446 &     2.1597 &      4.00 &      1.39 &      0.35 &      1.02 &      0.07 &     15.65 &      3.16 &      2.89 &      3.82 &       17.46$\pm$1.12 \\

  480666 &   149.4872 &     2.0303 &      4.18 &      0.60 &      0.14 &      0.24 &      0.07 &      3.60 &      0.14 &      2.91 &      9.40 &        7.45$\pm$2.07 \\

  315797 &   149.9304 &     1.7687 &      4.64 &      3.84 &      0.35 &      2.49 &      0.06 &     39.14 &      0.84 &      3.05 &      6.35 &       47.03$\pm$1.20 \\

  \\ 
\enddata \label{highz}
\tablenotetext{a}{The photometric redshifts and stellar masses of the galaxies are from \cite{ilb13} and the accuracy of those quantities is discussed in detail there. The SFRs are derived from the rest frame UV continuum and the infrared using Herschel PACS and SPIRE data as detailed in \cite{sco13}. All of the galaxies have greater than 10$\sigma$ photometry measurements so the uncertainties in M$_*$ and SFR associated with their measurements are always less than 10\%. As discussed in \cite{ilb13} the uncertainties in models used to 
derive the M$_*$ and SFR from the photometry are larger but generally less than a factor 2.}
\tablenotetext{b}{SNR$_{tot}$ and SNR$_{peak}$ are calculated separately and the column SNR lists the larger in absolute magnitude of those two SNRs. Note that we let the SNR be negative in cases where the flux estimate is negative so that several sigma negative flux values don't end up with a positive SNR above the detection thresholds.  }
\end{deluxetable*}

%\vskip -5in

 \begin{figure*}[ht]
\epsscale{1.}  
\plotone{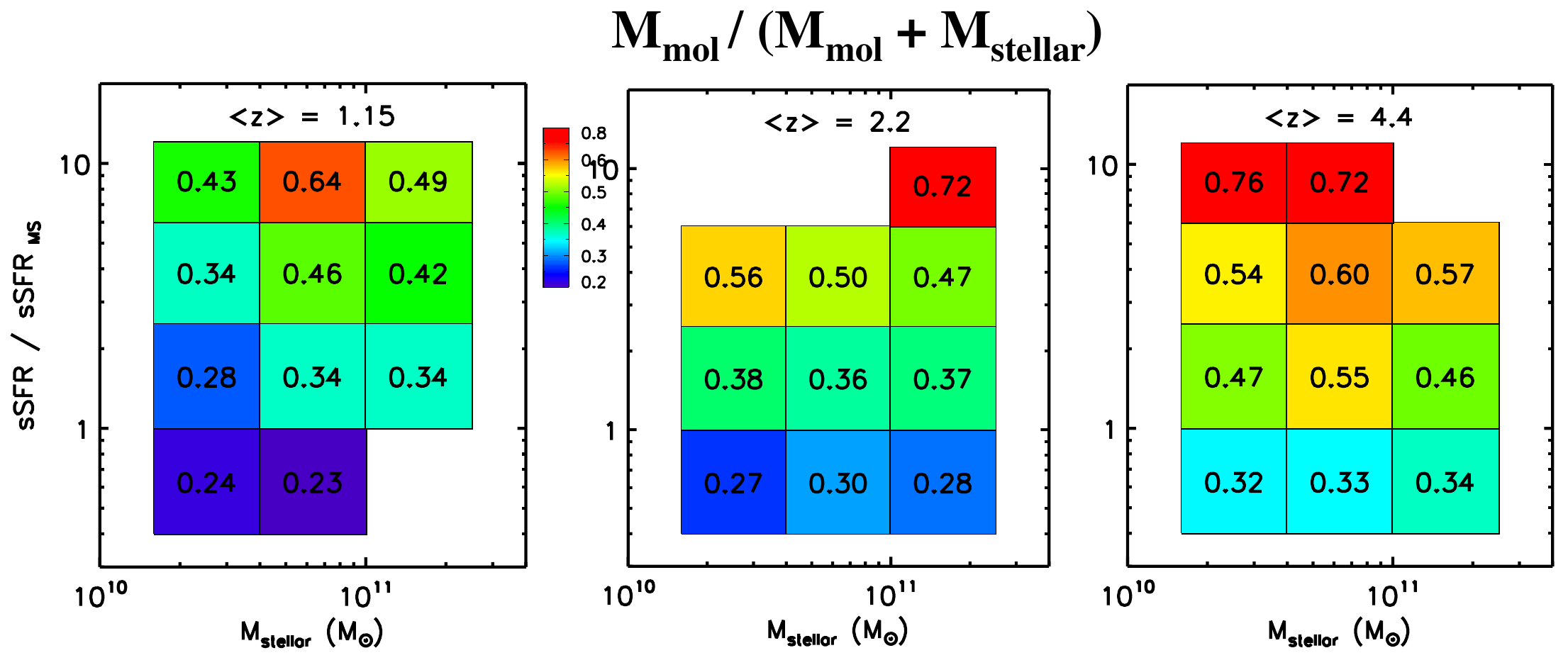}
\caption{Similar to Fig. \ref{stack_results} - bottom panel, except that here we show the values obtained from Equation \ref{gas-fraction_law} for the same M$_{stellar}$ and sSFR as the observed cells in Fig. \ref{stack_results}. The simple analytic fit in Eq. \ref{gas-fraction_law} adequately describes the trends in the observed cells
as a function of SFR, M$_{stellar}$ and redshift.}
\label{functions} 
\end{figure*}

\end{document}